\documentclass[a4paper,fleqn,usenatbib,useAMS]{mnras}

\usepackage{graphicx}	
\usepackage{amsmath}	
\usepackage{amssymb}	
\usepackage{multicol}        
\usepackage{bm}		
\usepackage{pdflscape}	

\usepackage{hyperref}
\usepackage[T1]{fontenc}
\usepackage{ae,aecompl}
\usepackage{newtxtext,newtxmath}

\title[$r$-process mixing]{Turbulent mixing of $r$-process elements in the Milky Way}

\author[P. Beniamini $\&$ K. Hotokezaka]{
	Paz Beniamini,$^{1}$\thanks{E-mail: paz.beniamini@gmail.com}
	Kenta Hotokezaka$^{2}$
	\\
	$^{1}$Division of Physics, Mathematics and Astronomy, California Institute of Technology, Pasadena, CA 91125, USA\\	
	$^2$Department of Astrophysical Sciences, Princeton University, 4 Ivy Lane, Princeton, NJ 08544, USA}
\pubyear{2020}
\begin{document}
\label{firstpage}
\pagerange{\pageref{firstpage}--\pageref{lastpage}}
\maketitle

\begin{abstract}
We study turbulent gas diffusion affects on $r$-process abundances in Milky Way stars, by a combination of an analytical approach and a Monte Carlo simulation. Higher $r$-process event rates and faster diffusion, lead to more efficient mixing corresponding to a reduced scatter of $r$-process abundances and causing $r$-process enriched stars to start appearing at lower metallicities. We use three independent observations to constrain the model parameters: (i) the scatter of radioactively stable $r$-process element abundances, (ii) the largest $r$-process enrichment values observed in any solar neighborhood stars and (iii) the isotope abundance ratios of different radioactive $r$-process elements ($^{244}$Pu/$^{238}$U and $^{247}$Cm/$^{238}$U) at the early solar system as compared to their formation.
Our results indicate that the Galactic $r$-process rate and the diffusion coefficient are respectively $r<4\times 10^{-5}\mbox{ yr}^{-1}, D>0.1 \mbox{ kpc}^2\mbox{Gyr}^{-1}$ ($r<4\times 10^{-6}\mbox{ yr}^{-1}, D>0.5 \mbox{ kpc}^2\mbox{Gyr}^{-1}$ for collapsars or similarly prolific $r$-process sources) with allowed values satisfying an approximate anti-correlation such that $D\approx r^{-2/3}$, implying that the time between two $r$-process events that enrich the same location in the Galaxy, is $\tau_{\rm mix}\approx 100-200\mbox{ Myr}$. This suggests that a fraction of $\sim 0.8$ ($\sim 0.5$) of the observed $^{247}$Cm ($^{244}$Pu) abundance is dominated by one $r$-process event in the early solar system. Radioactively stable element abundances are dominated by contributions from $\sim 10$ different events in the early solar system. For metal poor stars (with [Fe/H]$\lesssim -2$), their $r$-process abundances are dominated by either a single or several events, depending on the star formation history. 
\end{abstract}




\section{Introduction}
The nature of $r$-process producing sites in the Universe has been long debated \citep{burbidge1957RvMP,cameron1957,Lattimer1974}.
Over the last years, several independent lines of evidence have emerged in support of the dominant source of $r$-process events being rare (e.g. in comparison to regular core-collapse supernovae, henceforth ccSNe). This evidence stems from abundance patterns of dwarf galaxies \citep{Tsujimoto2014,ji2016Nature,Roederer2016,Beniamini2016}, from $r$-process elements observed in extremely metal poor stars \citep{vandevoort2015MNRAS,ishimaru2015ApJ,Shen2015,Beniamini2018,Naiman2018,Macias2019}, from the  abundance of radioactive isotopes, such as $^{244}$Pu, in the early and present solar system \citep{Ellis1996ApJ,Wasserburg1996ApJ,Fry2015ApJ,Hotokezaka2015,Lugaro2018,Bartos2019Natur,cote2019ApJb}, and most recently and directly from the observation of radioactive-heating powered kilonova accompanying the gravitational wave (GW) detected binary neutron star (BNS) merger, GW 170817 \citep{Kasliwal2017,Nakar2019}. The latter has shown that BNS mergers (that are much rarer than ccSNe) could produce copious amounts of $r$-process, which are consistent with the rate and $r$-process mass per event that are required by the other lines of evidence \citep{HBP2018}. At the same time, the association of BNS mergers as the major source of $r$-process elements encounters two potential obstacles relating  to the delays between binary formation and merger. First, depending on the shortest delay time, it may be difficult to explain the rapid production of $r$-process elements seen in the extremely metal poor stars \citep{Argast2004,Matteucci2014,ishimaru2015ApJ}.
Second, mergers with a long delay time may struggle to explain the late time decline of [Eu/Fe] as a function of [Fe/H] seen across Galactic stars \citep{HBP2018,cote2019ApJ,Simonetti2019}. The latter observation has been one of the arguments in support of the $r$-process production being dominated instead by collapsars \citep{Siegel2019}. Clearly, developing a better understanding of the expected abundance evolution can hold valuable information for deciphering the nature of the sources.

So far, studies have been mainly focused on the average trend of the $r$-process abundance evolution. The former can be well reproduced using a `one zone' treatment, in which the $r$-process event location does not matter and the synthesized material is assumed to be fully mixed in the Galaxy on a relatively short timescale, such that the only the time of occurrence of each event affects the overall evolution. An independent test involves considering also the fluctuations in the abundance evolution. This adds complexity to the problem, and requires a departure from a simple `one zone' model. In particular, the location of the events, as well as the turbulent diffusion of the $r$-process enriched gas within the Galaxy, have to be taken into account.
\cite{Argast2004} first developed a model taking into account inhomogeneity of $r$-process enrichment (see \citealt{Wehmeyer2015MNRAS,Cescutti2015} for recent studies). Although this model captures the spatial fluctuation of chemical abundances, it is assumed that each explosion enriches the surrounding gas until the end of the Sedov-Taylor phase, i.e., the turbulent diffusion is not taken into account. For stable isotopes, hydrodynamic simulations for galaxy evolution are in principle capable of treating both the spatial fluctuations and turbulent diffusion \citep{vandevoort2015MNRAS,Shen2015,Hirai2015ApJ,Safarzadeh2017MNRAS,Naiman2018}. For radioactive isotopes, on the other hand, hydrodynamic simulations must resolve the time scale of their half lives (see, e.g., \citealt{Vasileiadis2013ApJ,Fujimoto2018MNRAS,Krause2018A&A} for the mixing of $^{26}$Al and $^{60}$Fe).
Given the high computational cost of such simulations, it is useful to develop an analytic method taking the turbulent diffusion effect into account and providing a complementary treatment of the problem \citep{Hotokezaka2015,Krumholz2018MNRAS}. Such  analytic methods enable us to carry out a systematic study of a large parameter space as well as to evaluate the mixing of radioactive isotopes with different half-lives extending from a few Myr to Gyr. 
In this paper we focus on the latter. We constrain the rate of $r$-process events and the diffusion coefficient for turbulent gas diffusion, by (i) studying their affects on the scatter of stable $r$-process elements at a given value of [Fe/H], (ii) considering the upper limits on the rate of extremely enriched stars and (iii)  studying isotope abundance ratios of elements with different radioactive decay times.

\section{Model description}
\label{sec:model}
We construct a Monte Carlo simulation to calculate the effects of gas diffusion on the Galactic rare element abundances. For concreteness we focus here on $r$-process elements.
We begin by estimating the distances from the solar system and times at which the $r$-process events took place. For the distances we assume a distribution following that of the thin disk $\propto \exp[-z/h_z-(r-r_0)/h_r]$ where $z$ is the height above the plane, $r$ is the Galactic radius, $r_0=8\mbox {kpc}$ is the distance of the solar system from Galactic center and the scale heights are $h_z=0.25\mbox{ kpc}$ and $h_r=3.5\mbox{ kpc}$ \citep{BS1980}. Later on, in \S \ref{sec:varassump}, we consider also the possibility that $r$-process events take place at a significant offset relative to the disk.
The distribution for the time of each $r$-process event is given by the convolution of the star formation rate (SFR), $\Psi(t)$, and the delay time distribution (DTD), $DTD(t)$. The probability of an event per unit time is then
\begin{equation}
    \frac{dP}{dt}\propto \int_0^t DTD(t-t') \Psi(t') dt'.
\end{equation}
We denote the resulting rate of $r$-process events (proportional to $dP/dt$ above) as $r(t)$.
We assume the star formation follows either the Cosmic SFR \citep{Madau2014} or a constant rate. The latter has been considered as a good approximation for the Milky Way (MW) SFR history based on the ages of nearby late-type dwarfs (\citealt{Rocha-Pinto2000A&A} see also \citealt{Tremblay2014ApJ} based on  white dwarfs). For collapsars or other ccSNe, a very short delay time is expected between the formation of the stellar progenitors and core-collapse. For BNS mergers, the DTD has been recently estimated by \cite{BP2019} using the observed Galactic BNS systems. These authors find that a DTD significantly steeper than $t^{-1}$ (i.e. more short delays) is required by the data. Regardless of the dominant $r$-process sources, short delays for $r$-process production, or a steeper DTD than $t^{-1}$, are also required to explain the rapid observed decline of [Eu/Fe] as a function of [Fe/H] at $\mbox{[Fe/H]}>-1$ \citep{HBP2018,cote2019ApJ,Simonetti2019} and the observed $r$-process abundances of ultra-faint dwarf galaxies \citep{BHP2016}. We therefore assume the case of short delay times as our fiducial case \footnote{An absolute minimum on the delay time arises from the life time of the progenitor stars. However, since the stars in question are massive, this minimum delay is short relative to the typical timescales of evolution for the star formation and the mean iron abundances (which at the early solar system are of order Gyrs).}. However, for the sake of completeness we return to address the role of the DTD in \S \ref{sec:varassump} where we assume the commonly employed delay time, $\propto t^{-1}$ and examine its affects on the analysis presented here. The true delay time is expected to lie somewhere between these two extremes.

The production of iron is assumed to be dominated by core-collapse, and at later times, Ia supernovae. The evolution of the iron abundance as a function of time is calculated according to the one zone model described in \citep{HBP2018}, and with the same model parameters (see table 1 of \citealt{HBP2018} for a list of the model parameters). The underlying assumption is that due to its much higher production rate, the iron abundance progresses smoothly with time and fluctuates much less than the $r$-process abundances. The validity of this approximation and the implications of relaxing this assumption are explored in detail in \S \ref{sec:ironstochastic}.

To calculate the $r$-process abundance of stars born at a given time we take into account contributions from all the events that took place up to a time $t_{\rm *}$ before the star was born (where $t_*\approx 10^{7}$ yr is the typical time it takes to reprocess gas into stars. This is comparable to an upper limit of $T_{\rm isolation}$ as defined in \citealt{Lugaro2018}).
At a given time and location $t,\vec{r}$ the density of a radioactive element is given by the contribution of all past events, taking into account their (possibly) finite radioactive lifetimes and the turbulent diffusion of the enriched gas in the MW disk (see also \citealt{Hotokezaka2015}),
\begin{equation}
\label{eq:sum}
    \rho(t,\vec{r})=\sum_{t_j<t-t_*}\frac{m_r e^{-f(t) \Delta t_j}}{K_j(\Delta t_j)} \exp{\bigg[-\frac{|\vec{r}-\vec{r}_j|^2}{4D \Delta t_j}-\frac{\Delta t_j}{\tau}\bigg]}
\end{equation}
where the index $j$ sums over the historic events (with locations and times $\vec{r}_j, t_j$ correspondingly), $m_r$ is the mass of the element produced per event, $f(t)$ is the mass-loss rate of the element in the interstellar medium (ISM) due to star formation and galactic outflows (see \citealt{HBP2018}), $D$ is the turbulent diffusion coefficient, $\tau$ is the radioactive lifetime of the element, $\Delta t_j=t-t_j$ and $K_j(\Delta t_j)$ is given by
\begin{equation}
    K_j(\Delta t_j)=\min[(4\pi D \Delta t_j)^{3/2},8\pi h_z D \Delta t_j].
\end{equation}
$K_j(\Delta t_j)$ thus encompasses the transition between spherical diffusion for close by events to planar diffusion for more distant ones. In addition, once the diffusion length becomes larger than the Galactic radius ($\sqrt{D \Delta t_j}\gtrsim 4.6 h_r$) we limit $K_j(\Delta t_j)$ to the Galactic volume.

The mixing process of metals injected by an astrophysical event in the ISM proceeds in the following way. After the injection, metals expand as a blast wave until the velocity of the wave becomes comparable to the ISM sound speed, which occurs on a timescale of a few Myr for supernovae and neutron star mergers (e.g. \citealt{Haid2016,Beniamini2018}).
Then, metals diffuse with the turbulent mixing process, of which the diffusion coefficient may be described by
\begin{equation}
D = \alpha c_s H \approx \alpha\ \left(\frac{c_s}{10\,{\rm km/s}} \right)
\left(\frac{H}{100\,{\rm pc}} \right)
{\rm kpc^2Gyr^{-1}},
\end{equation}
where $c_s$ is the ISM sound speed, $H$ is the scale height of the ISM, and $\alpha$ is mixing parameter.
A precise apriori determination of $\alpha$ would require knowledge of the specific physical processes that drive the turbulent diffusion. The advantage of the $\alpha$ formulation is that it encapsulates our ignorance into the dimensionless parameter $\alpha$ and provides an approximate scale for the expected value of $D$. For this reason we do not assume $D$ to be a known quantity in this work, and instead find constraints on its value directly from observations.
Since the blast-wave propagation timescale is short relative to the other timescales involved, most of the mixing proceeds through the turbulent diffusion.

It is useful to define a characteristic time,
$\tau_{\rm mix}$, that is the typical time between two events enriching the same location through turbulent diffusion. $\tau_{\rm mix}$ can be approximated by \citep{Hotokezaka2015}:
\begin{equation}
\label{eq:nratio}
    \tau_{\rm mix}=200 \bigg(\frac{r}{3\times 10^{-5}\mbox{ yr}^{-1}}\bigg)^{-2/5}\bigg(\frac{D}{0.1 \mbox{ kpc}^2\mbox{ Gyr}^{-1}}\bigg)^{-3/5}\mbox{ Myr}.
\end{equation}
An additional source of mixing (that we do not treat in this work) may be provided by the differential rotation of the Galaxy. Differential rotation will generally become important on longer timescales. We estimate the characteristic time-scale for this process in the following way. Consider the difference in propagation in the azimuthal direction, $\Delta r_{\phi}$, between two points that are separated in the radial direction by a distance of one diffusion length $\Delta r=\sqrt{Dt}$,
\begin{equation}
  \Delta r_{\phi}=\int \bigg(r \frac{d\Omega}{dr}\bigg)_{R} \Delta r dt \approx  \bigg(r \frac{d\Omega}{dr}\bigg)_{R} D^{1/2} \frac{2}{3}t^{3/2}
\end{equation}
where $\bigg(r \frac{d\Omega}{dr}\bigg)_{R}$ is the rotational velocity gradient at $R$, the radius of the solar system from the Galactic center, and is $\bigg(r \frac{d\Omega}{dr}\bigg)_{R}\approx 10 \mbox{km s}^{-1}\mbox{ kpc}^{-1}$ \citep{Roy1995}. Differential rotation will become important if $\Delta r_{\phi}\gtrsim few \Delta r$. This corresponds to a critical timescale of
\begin{equation}
    \tau_{\rm rot}\approx 3\times \frac{3}{2} \bigg(r \frac{d\Omega}{dr}\bigg)_{R}^{-1}\approx 450\mbox{ Myr}.
\end{equation}
This suggests that if the contribution to the enrichment of a given element is dominated by events taking place more than $\sim 0.5$ Gyr before star formation, then differential rotation can no longer be ignored as it will become an important source of mixing. As we will show below, this is not a major concern for the parameter space of interest found in our analysis (see \S \ref{sec:varassump} for more details).

For each Monte Carlo simulation we record the median and scatter of various abundances at the ESS.
The Monte Carlo calculation is then repeated 100 times (each time drawing enrichment locations and times according to the underlying distributions and evolving the model until the time of the ESS), each time yielding a different abundance evolution. Comparing the different simulation runs we then obtain the typical values of the parameters of interest and the error in their values that arise due to the inherently stochastic nature of the process.

\section{Results}
\subsection{The scatter of element abundances} 
\label{sec:scatter}
The observed scatter in element abundances is related to the rate of the events leading to their formation, $r$, and to the turbulent diffusion coefficient of gas in the Galaxy, $D$. Faster diffusion and higher formation rates both lead to more efficient mixing of the gas and thus reduce the scatter. Therefore, a relation between $r,D$ can be established by comparing the simulated results with observations.

\begin{figure}
\centering
\includegraphics[width=0.45\textwidth]{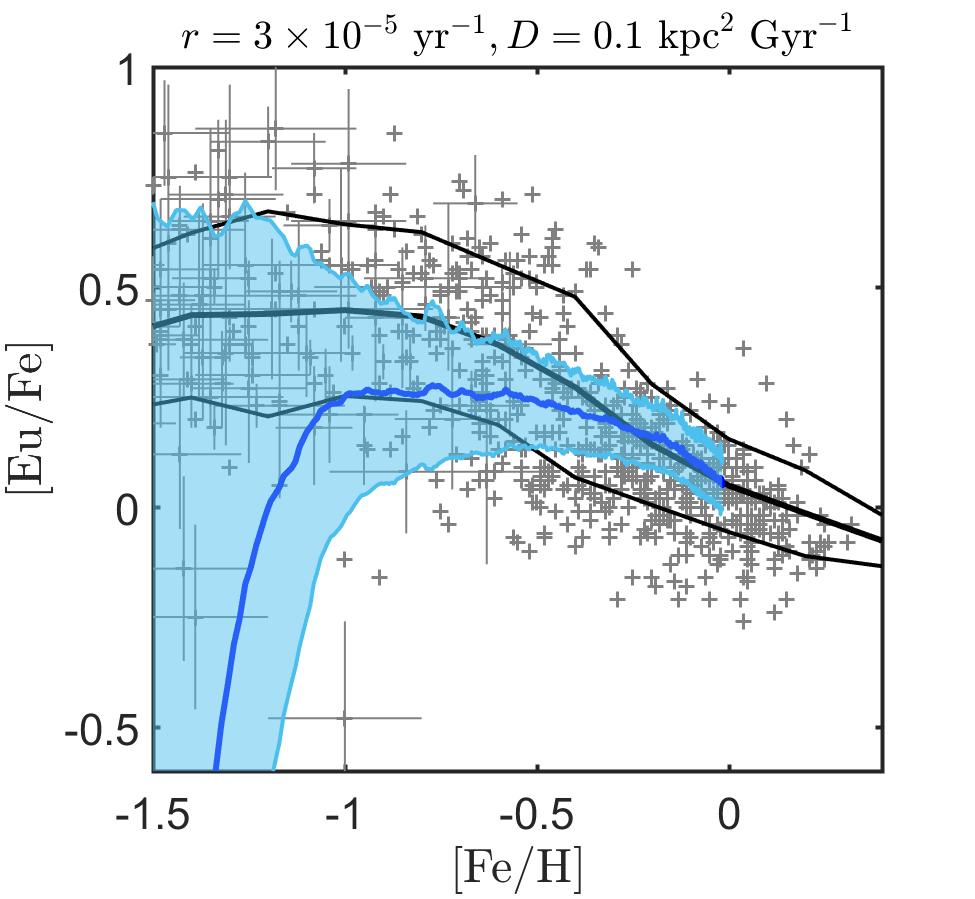}\\
\includegraphics[width=0.45\textwidth]{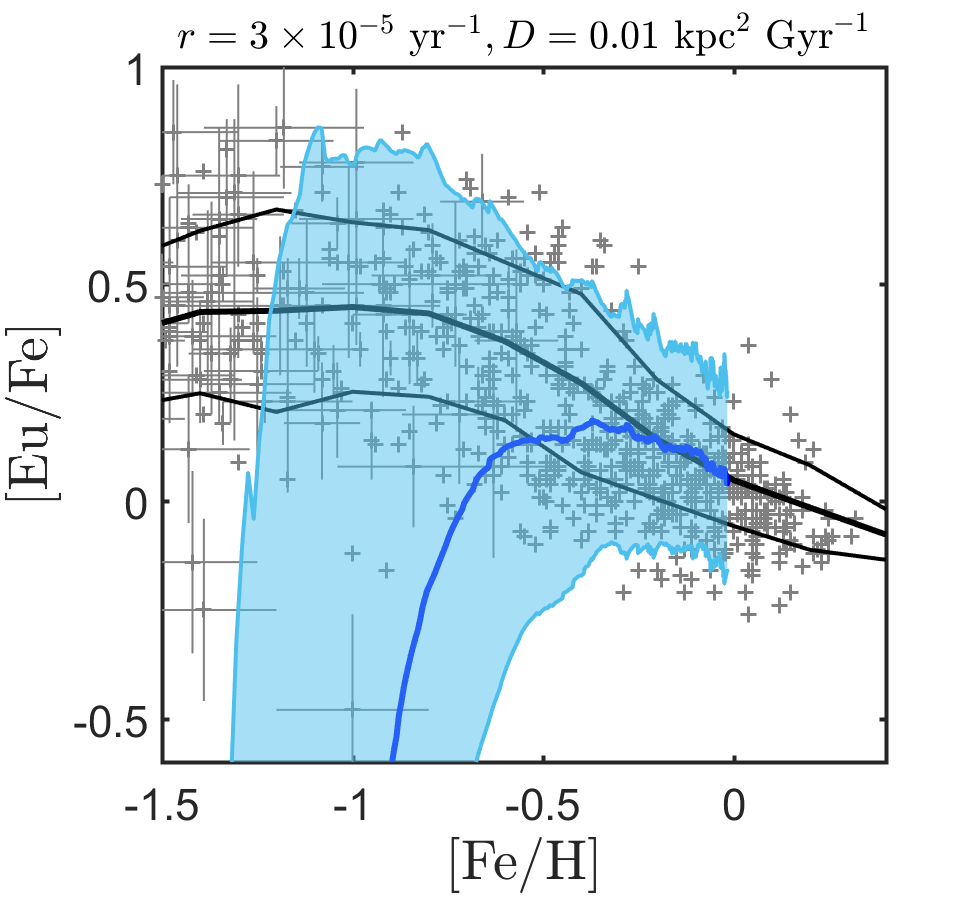}
\caption{Average trend and 1$\sigma$ scatter in [Eu/Fe] as a function of [Fe/H] for observed MW stars (black lines) and for a Monte Carlo simulation with a constant star formation rate, no delay between star formation and the $r$-process forming events, an $r$-process rate of $r=3\times 10^{-5}\mbox{ yr}^{-1}$ and diffusion coefficients of either $D=0.1 \mbox{ kpc}^2\mbox{ Gyr}^{-1}$ or $D=0.01 \mbox{ kpc}^2\mbox{ Gyr}^{-1}$.}\label{fig:EuFescatter}
\end{figure}

In figure \ref{fig:EuFescatter} we present a comparison of the simulated results with observed europium abundances taken from the SAGA database \citep{Suda2008PASJ}, where the abundance measurements from different observations are compiled (see \citealt{Battistini2016A&A,Griffith2019ApJ} for homogeneous data set). For this figure we have adopted a constant formation rate and no delay between star formation and the following $r$-process events. We focus here on the Eu abundance, since this element is known to be dominated by the $r$-process and can be easily compared with observations. We also focus on $\mbox{[Fe/H]}\ge -1.5$, since at lower metallicites, the stars could be predominantly originating from dwarf galaxies that have been tidally stripped onto the MW halo during the halo assembly process (see \citealt{Beniamini2018} and references therein).
As is seen in the figure, a Galactic rate of $r=3\times 10^{-5}\mbox{ yr}^{-1}$ (typical for BNS merger estimates) and a diffusion coefficient of $D\approx 0.1 \mbox{ kpc}^2\mbox{ Gyr}^{-1}$ (corresponding to $\alpha$ disc models for the MW with $\alpha\approx 0.1$) can mimic reasonably well the observed scatter, even with little intrinsic scatter in the amount of $r$-process material formed in different events. Decreasing either the rate or the diffusion coefficient, broadens the abundance curves, and causes the [Eu/Fe] curve to rise at a later time and hence at a larger value of [Fe/H]. This is a reflection of the fact that the effective time between two $r$-process events enriching a given location, $\tau_{\rm mix}$, grows longer (see equation \ref{eq:nratio}). This is also the fundamental reason that we find a negative correlation between the required values of $D, r$ in our analysis detailed below.

\begin{figure}
\centering
\includegraphics[width=0.45\textwidth]{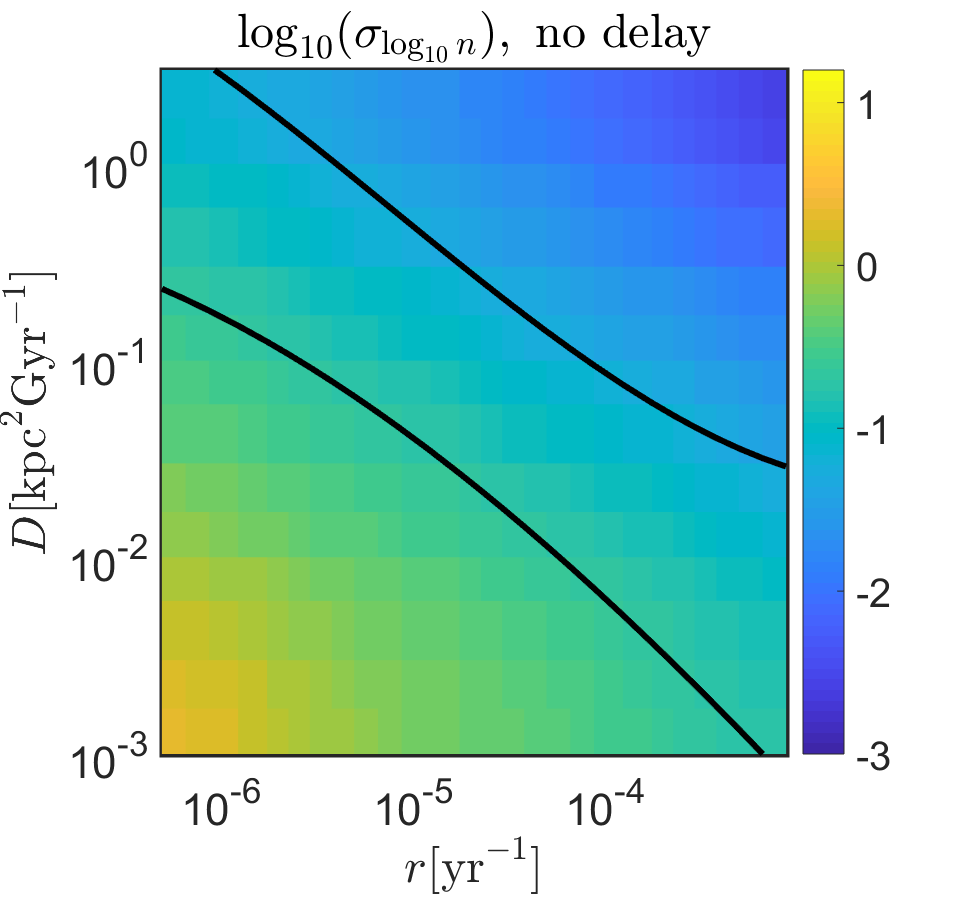}
\caption{standard deviation in the density of a stable element at the time of the solar system formation ($\log_{10}[\sigma_{\log_{10}(n)}]$) as a function of the $r$-process Galactic event rate, $r$, and the diffusion coefficient, $D$. The region that results in the same scatter as seen in observations of MW stars is between the two black lines. In particular, the regime below the bottom line is strongly ruled out by our analysis.}\label{fig:scatterregion}
\end{figure}

Generalizing these results, we present in figure \ref{fig:scatterregion} the allowed region in the $r$-$D$ parameter space in which the simulated mixing scatter is consistent with the observed one. We note that although it is possible to account for some of the observed scatter with effects other than the mixing scatter (e.g. varying the intrinsic $r$-process mass per event), any additional such source of variation is unlikely to be correlated with the mixing scatter presented here in such a way as to {\it reduce} the observed scatter. Therefore, the region below the bottom black line in figure \ref{fig:scatterregion} is strongly ruled out by this consideration.

The overall scatter is a good measure of the abundance differences that may be seen between different MW stars. There is yet more information embedded in the spatial distribution of these variations, provided that one can make a secure connection between the present day separation of stars and their separation at birth and assuming that one can identify with confidence pairs of stars that were born at roughly the same time (and with similar metalicity). From equation \ref{eq:sum} it is clear that, given these conditions, stars close to the sun, should exhibit very similar abundances to the sun (the temporal and distance of all enrichment events are similar for them and for the sun), while stars that are further away will gradually show larger fluctuations. This trend will eventually saturate at the global level of abundance fluctuations (which is shown in figure \ref{fig:scatterregion}). In particular, this consideration leads to a lower limit on the diffusion coefficient based on the small variation in the chemical abundances of nearby open clusters.  \cite{Friel1992ApJ} measure the chemical abundance of four open clusters younger than $\sim 4\cdot 10^8$ yrs, for which cluster-to-cluster variations are consistent with zero within the measurement error. Given that the separation between these clusters is $\sim 100$ pc, we obtain a lower limit of the diffusion coefficient,  $D\gtrsim 0.01\,{\rm kpc^2/Gyr}$.

\begin{figure}
\centering
\includegraphics[width=0.4\textwidth]{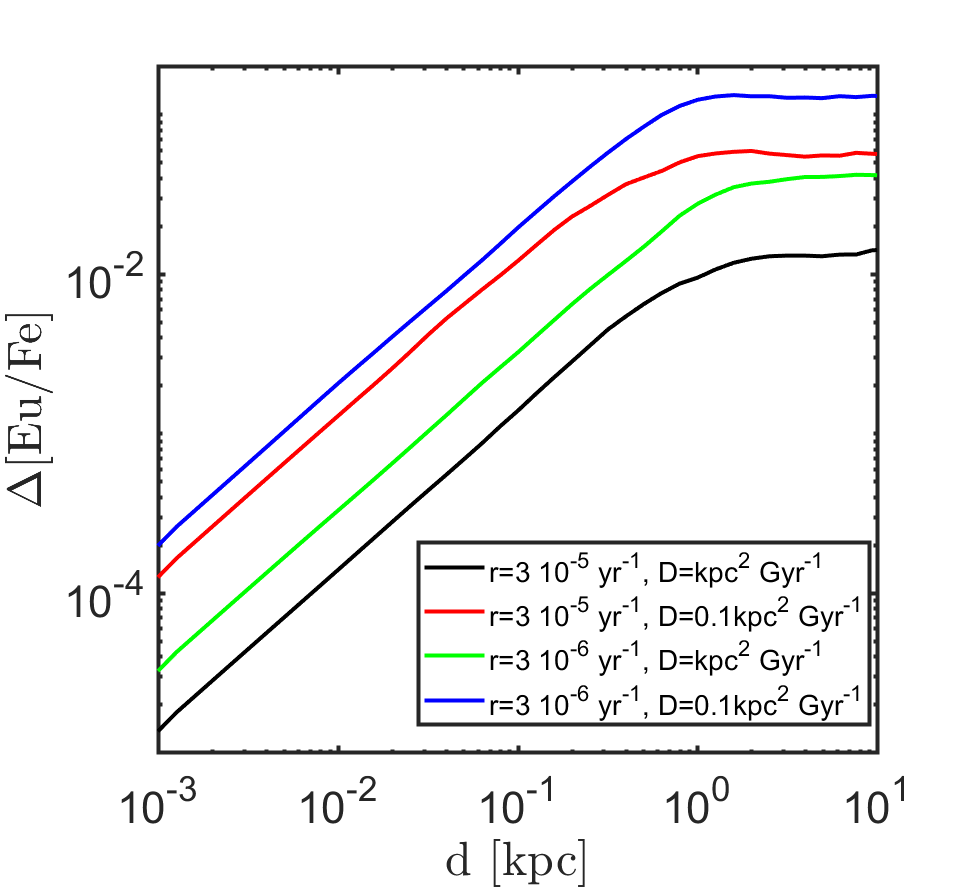}
\caption{Fluctuations in the abundance of a stable $r$-process element (e.g. [Eu/Fe]) between stars born at the same time (with the same [Fe/H]) and at a distance $d$ between them. Results are shown for different $r$-process rate and diffusion coefficients.}\label{fig:spatialscatter}
\end{figure}

The spatial dependence of the abundance fluctuations is shown in figure \ref{fig:spatialscatter}, in which we consider stars born at the same time (with the same iron abundance) but with a distance $d$ between them. For a given rate and diffusion coefficient we repeat our Monte Carlo simulation, described in \S \ref{sec:model}, 200 times. We record the median level of stable $r$-process element abundances between pairs of stars born at the same time at different locations. Larger rates and / or diffusion coefficients (corresponding to more efficient mixing), result in smaller fluctuations at any distance. We also see that up to a distance of $\sim$ kpc, the star to star fluctuations remain smaller than the global level of fluctuations. At smaller distances, the fluctuations in [Eu/Fe] are approximately linear with distance (or equivalently, the fluctuations in the $r$-process density increase roughly exponentially with distance). This can be easily understood in the following way. Consider an enrichment event at a location $\vec{l}_1$ from star 1 ($\vec{l}_2$ from star 2). The location of star 2 relative to 1 is $\vec{d}=\vec{l}_1-\vec{l}_2$.
From equation \ref{eq:sum}, the difference in the contribution to the $r$-process density from this single event is
\begin{equation}
    \frac{\rho_1}{\rho_2}=\exp\bigg[ -\frac{|\vec{l}_1|^2}{4D\Delta t}+\frac{|\vec{l}_1-\vec{d}|^2}{4D\Delta t} \bigg] \approx \exp\bigg[ \frac{l_1d}{5.6D\Delta t}\bigg]
\end{equation}
where in the R.H.S. we have taken star 1 to be the more enriched of the two and have expanded the expression to first order in $d/l_1$. In particular, since the distance of a typical enrichment event scales as $l_1\propto \sqrt{D\tau_{\rm mix}}$, we obtain that the pre-factor of $d$ scales as $r^{-1/5}D^{-4/5}$. These scalings are reproduced in the numerical calculation in figure \ref{fig:spatialscatter}. 
Strictly speaking, the spatial dependence discussed here applies to stars with the same Galactocentric radii. When considering stars that are radially separated, differential rotation could become important once the distance becomes $d\gtrsim \sqrt{D\tau_{\rm rot}}$, which is roughly $0.2\mbox{ kpc}$ for $D=0.1\mbox{ kpc}^2\mbox{ Gyr}^{-1}$. Furthermore, stars that have migrated significantly since their birth, may show larger fluctuations at a given distance than expected from the above estimate. This would affect mostly older stars and those with larger proper motions.

\subsection{Extremely enriched stars}
\label{sec:rare}
The abundance patterns are informative not only in terms of the median and scatter of the abundance ratios, but also in terms of the expectations for particularly rare events. A small fraction of the observed stars should have been formed physically and temporally close enough to an enrichment event that they would exhibit extreme levels of enrichment. Comparing the expectations for these enrichment levels with the maximal abundance levels in any of the observed stars puts therefore a limit on the rate and / or the Diffusion coefficient for turbulent mixing.

The Galactic $r$-process rate considered in this paper, $r$, can be converted to a local (within the solar system neighborhood) volumetric rate of
\begin{equation}
    R=1.4\times 10^{-10} \bigg(\frac{r}{3\times 10^{-5}\mbox{ yr}^{-1}}\bigg) \mbox{Myr}^{-1}\mbox{ pc}^{-3}
\end{equation}
Consider an $r$-process enrichment event that occurs at a time interval $T$ before a certain cloud of gas begins to form stars. The probability that an event has occurred within one diffusion length: $l=\sqrt{DT}$ and up to a time $T$ before star formation, is simply
\begin{equation}
\label{eq:P}
    P\!\sim \! RT\frac{4\pi}{3} l^3\!=\!3\times 10^{-3} \bigg(\frac{r}{3\times 10^{-5}\mbox{ yr}^{-1}}\bigg) \bigg(\frac{D}{0.1 \mbox{ kpc}^2\mbox{ Gyr}^{-1}}\bigg)^{3/2}  T_{30}^{5/2}  
\end{equation}
where $T_{30}\equiv \bigg(\frac{T}{30\mbox{ Myr}}\bigg)$.
For our fiducial rate and diffusion coefficient, the number of stars with measured abundances, $N_{\rm star}\approx 10^3$ is such that $N_{\rm star}\gg P^{-1}$, implying that there should be a sub-sample of the observed stars which have had an enrichment event within a time $T$ (and distance $l(T)$) in their past.

With $N_{\rm star}\sim 10^3$, the $2\sigma$ probability threshold for there being no observed stars with this level or higher of enrichment, translates to $P^{-1}\gtrsim N_{\rm star}/3$. For a given $D,r$, there is a maximum value of $T\equiv T_{\rm P}$ (corresponding to a lower limit on the enrichment of extremely $r$-process rich stars) for which this condition is satisfied. We shall consider values of $T$ such that $T=\max(T_{\rm P},T_{\rm fade})$. $T_{\rm fade}$ is the time it takes the enriched matter from a blast wave carrying $r$-process material to become incorporated into the interstellar medium. It is given by $T_{\rm fade}\approx 5 E_{51}^{0.32}n_{-1}^{-0.37}c_{\rm s,1}^{-7/5}\mbox{ Myr}$ \citep{Draine2011}, where $E_{51}\equiv E/10^{51}\mbox{ erg}, n_{-1}=n/0.1 \mbox{cm}^{-3}, c_{\rm s,1}=c_{\rm s} / 10 \mbox{ km s}^{-1}$. It represents a lower bound on the time-scale before turbulent mixing can set in and equation \ref{eq:P} may be applied. The time $T$ is therefore the minimum time between an $r$-process event and star formation for which it is statistically ensured that enrichment has occurred within $l(T)$ and has been well mixed into the ISM.

\begin{figure}
\centering
\includegraphics[width=0.45\textwidth]{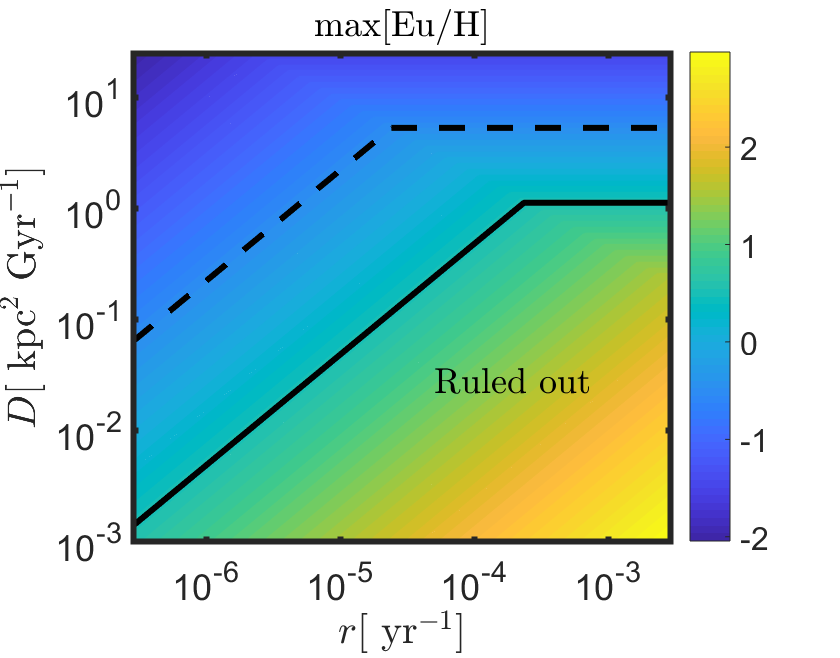}
\caption{Maximum expected level of [Eu/H] for $10^3$ solar neighborhood stars. Results are plotted for $n=0.3 \mbox{ cm}^{-3}, E=10^{51} \mbox{ erg}, c_{\rm s}=10\mbox{ km s}^{-1}$ and $m_{\rm Eu}=10^{-4}M_{\odot}$. Below the solid line, the results are in contrast with the observed value. Also shown in a dashed line, is the equivalent limit for $m_{\rm Eu}=10^{-3}M_{\odot}$ (and other parameters as before).}\label{fig:maxEuH}
\end{figure}

For a given $T,l(T)$, one may estimate the expected $r$-process enrichment relative to hydrogen. The density of a stable $r$-process material that has been spread over a distance $l(T)$ is given by equation \ref{eq:sum}, $\rho_r(T,l(T))=m_re^{-1/4}/(4\pi D T)^{3/2}$ (where $m_r$ is the mass of stable $r$-process produced in one event). Considering Eu and comparing to the hydrogen density we find
\begin{equation}
    \frac{\rho_{\rm Eu}}{\rho_{H}}=4.3\times 10^{-9} m_{\rm Eu,-4}n_{-1}^{-1}T_{30}^{-3/2}\bigg(\frac{D}{0.1 \mbox{ kpc}^2\mbox{ Gyr}^{-1}}\bigg)^{-3/2}
\end{equation}
where $m_{\rm Eu,-4}\equiv m_{\rm Eu}/10^{-4}M_{\odot}$.
This density ratio can be written in terms of an abundance, using $\mbox{[Eu/H]}=9.43+\log_{10}(\rho_{\rm Eu}/\rho_H)$. Using our fiducial parameters this yields $\mbox{[Eu/H]}\approx 1$ which is higher than the observed abundance of any of the stars in our sample. The latter (accounting for measurement errors) is $\mbox{[Eu/H]}_{\rm max}=0.5$. Requiring $\mbox{[Eu/H]}<\mbox{[Eu/H]}_{\rm max}$ therefore leads to a lower limit on the diffusion coefficient at a given rate
\begin{eqnarray}
  &  D\geq  \min\bigg[0.84 \bigg(\frac{r}{3\times 10^{-5}\mbox{ yr}^{-1}}\bigg) m_{\rm Eu,-4}^{5/3} n_{-1}^{-5/3},\\& 1.6 E_{51} m_{\rm Eu,-4}^{2/3} n_{-1}^{-0.3} c_{\rm s,1}^{7/5}\bigg] {\rm kpc^2Gyr^{-1}}
\end{eqnarray}
The allowed parameter space is presented in figure \ref{fig:maxEuH}. As a conservative approach we assume in that figure a relatively large value of the local ISM density, $n\approx 0.3 \mbox{ cm}^{-3}$. We find that a combination of a low rate and a large diffusion coefficient are required in order not to overproduce the maximal observed value of $r$-process enrichment.  The limits become significantly more constraining for larger $r$-process masses produced per event as can be seen from the dashed line in figure \ref{fig:maxEuH}.

\subsection{Isotope abundance ratios}
\label{sec:ratios}
It is instructive to consider not only stable isotopes, but also elements with different radioactive timescales. 
Since different elements have different decay times, the abundance ratio of different elements can be different to the production ratio and is sensitive to $r$ and $D$.

As an illustration, consider two elements: $A,B$ that are both dominated by a single type of event in their past.  When considering the abundance ratio of such radioactive elements in the ISM at a given location and time, there are two limits: (1) a stationary regime ($\tau_{A,B}\gg \tau_{\rm mix}$) and (2) a single event regime ($\tau_{A,B}\ll \tau_{\rm mix}$), where $\tau_{\rm mix}$ is the typical time between two enrichment events that enrich the ISM in the same location. In the stationary regime, multiple astrophysical events contribute to the enrichment and the abundance ratio is given by  
\begin{equation}
    \bigg<\frac{n_B}{n_A}\bigg>\approx \bigg(\frac{n_{B}}{n_{A}}\bigg)_0\frac{\tau_B}{\tau_A}\label{eq:st}
\end{equation}
where $\bigg(\frac{n_{B}}{n_{A}}\bigg)_0$ is the abundance ratio of the two elements at the production phase, $\tau_A, \tau_B$ are the radioactive mean-lives of elements $A,B$ respectively. 

In the single event regime, the median ratio (averaging over the time of observation relative to the enrichment times) of their abundances at some later point in the future is expected to be \footnote{The expression represents the median value of said measurement as well as the typical deviation (up to a factor of order unity) around this value.}
\begin{equation}
    \bigg<\frac{n_B}{n_A}\bigg>\approx \bigg(\frac{n_{B}}{n_{A}}\bigg)_0\exp\bigg[-\frac{\tau_{\rm mix}}{2} \bigg(\frac{1}{\tau_B}-\frac{1}{\tau_A}\bigg)\bigg]\label{eq:single}
\end{equation}
Since $\tau_{\rm mix}$ is a function of $r$ and $D$, measuring the abundance ratio of two elements at a certain point in time, compared to their production ratio, can be used to constrain $r,D$.

We now apply this idea to the radioactive $r$-process elements: $^{238}$U, $^{244}$Pu, and $^{247}$Cm, of which the mean-lives are   $\tau_{^{238}{\rm U}}=6.4\mbox{ Gyr}$, $\tau_{^{244}{\rm Pu}}=117\mbox{ Myr}$, $\tau_{^{247}{\rm Cm}}=22.5\mbox{ Myr}$  \citep{ENDF}. Note that these isotopes are particularly important because of the following reasons: (i) they are purely $r$-process elements, (ii) their abundance ratios at the early solar system (ESS) are measured, and (iii) $\tau_{\rm mix}$ is expected to be between $\tau_{^{247}{\rm Cm}}$ and $\tau_{^{238}{\rm U}}$.

The abundance ratio of $^{244}$Pu/$^{238}$U and $^{247}$Cm/$^{238}$U of the ESS is $\approx 0.008$ and $\approx 2\cdot 10^{-5}$ \citep{Turner2007EPSL,Tissot2016Sci}, respectively, while the values expected from equation (\ref{eq:st}) are $0.02$ and $4\cdot 10^{-3}$, where we assume $\bigg(\frac{n_{B}}{n_{A}}\bigg)_0=1$ (this is comparable to calculations from nuclear physics networks  \citealt{Cowan1991PhR, Eichler2015ApJ} and Eichler. private communication). Thus, the stationary approximation overestimates the abundance ratios, suggesting that there was a significant time delay (relative to the radioactive decay scales) between the latest $r$-process event that enriched the local gas before the ESS and the ESS formation. Note also that live $^{244}$Pu particles from the ISM are currently accumulating on the Earth's deep sea floor and the measured flux is lower by at least factor of $10$ than the value expected from the stationary approximation \cite{Paul2001ApJ,Wallner2015}. These measurements indicate $\tau_{\rm mix}\gtrsim \tau_{\rm ^{244}Pu},\tau_{\rm ^{247}Cm}$.

The failure of the stationary approximation, motivates us to explore the opposite extreme, the single event regime. We apply equation (\ref{eq:single}) to the ESS abundance of $^{244}$Pu and $^{247}$Cm, and find $\tau_{\rm mix}\sim 330\,{\rm Myr}$ for $^{247}$Cm/$^{244}$Pu, which is longer than the mean-life of both elements. This suggests that the $^{244}$Pu and $^{247}$Cm contained in the ESS material are produced predominately by a r-process single event and the time separation between the ESS formation and the an r-process event is $\sim 330\,{\rm Myr}$ (see also \citealt{Tissot2016Sci,Bartos2019Natur,cote2019ApJb}). 
On the contrary, we obtain $\tau_{\rm mix}\sim 1.1\,{\rm Gyr}$ and $\tau_{\rm mix}\sim 490\,{\rm Myr}$ for the abundance ratios of $^{244}$Pu/$^{238}$U and $^{247}$Cm/$^{238}$U respectively. These values of $\tau_{\rm mix}$ are much shorter than the mean-life of $^{238}$U. The implication is that the event which predominantly enriched $^{244}$Pu and $^{247}$Cm of the ESS is not necessarily the main source of $^{238}$U. We will discuss the number of events contributing to enrichment in \S \ref{sec:Numberevents}.

\begin{figure}
\centering
\includegraphics[width=0.4\textwidth]{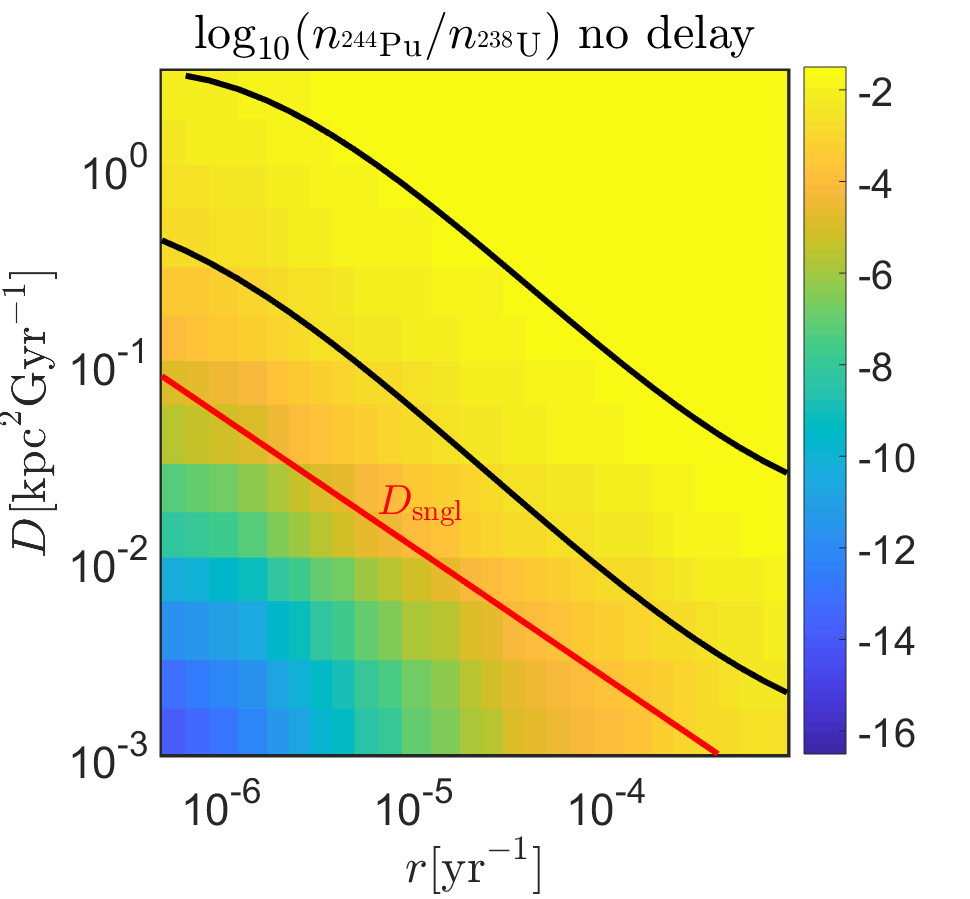}
\includegraphics[width=0.4\textwidth]{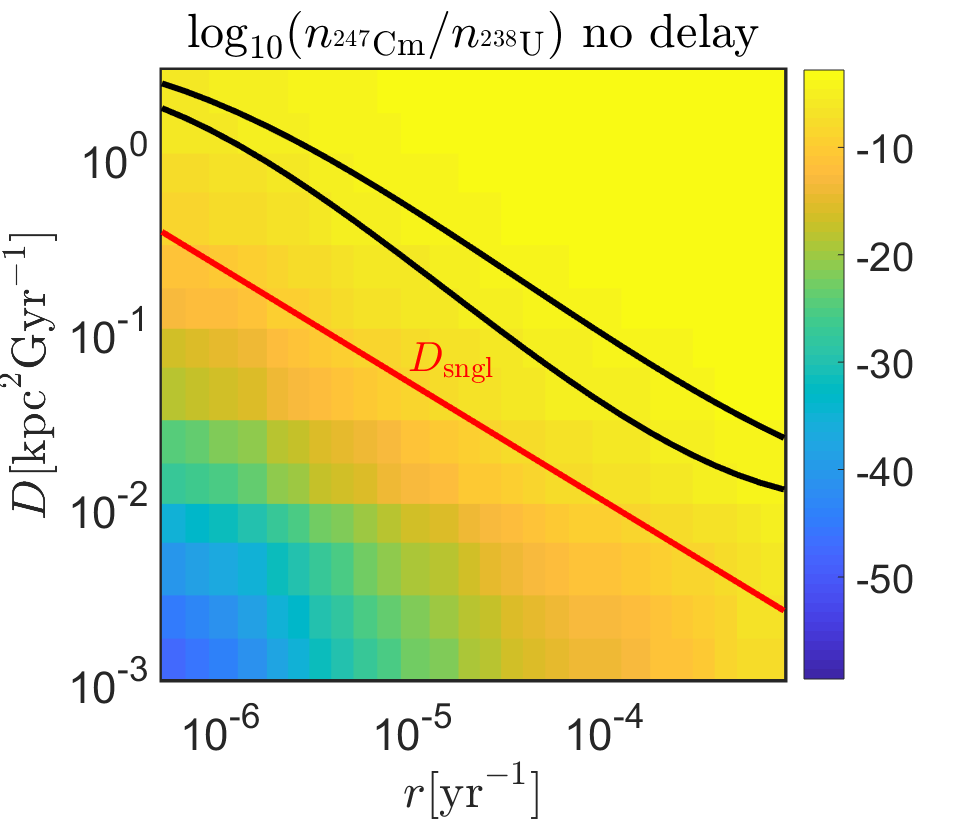}
\caption{Results of a Monte Carlo simulation for the ratio between the median value of different abundance ratios as measured at the ESS as compared to their production ratio. The region that is consistent with the observed values lies within the solid black lines, while the analytic estimate based on the assumption of a single enrichment event is depicted by a thick red line.}\label{fig:ratios}
\end{figure}

Plugging the single event regime estimate for $\tau_{\rm mix}$ back into equation \ref{eq:nratio} yields 
\begin{equation}
    D_{\rm sngl}\approx 0.0054-0.043 \bigg(\frac{r}{3\times 10^{-5}\mbox{ yr}^{-1}}\bigg)^{-2/3}\mbox{ kpc}^2 \mbox{ Gyr}^{-1}
\end{equation}
where the higher number is for the $^{247}$Cm/$^{244}$Pu ratio and the lower for the $^{244}{\rm Pu}/^{238}{\rm U}$ ratio (the value for $^{247}{\rm Cm}/^{238}{\rm U}$ is in between the two). This simple estimate can be compared with the results of the Monte Carlo simulation (where it is not assumed that the production of any of the elements are dominated by a single event or by the stationary approximation and where we do not average over the solar system formation time relative to the time of enrichment as done in deriving equation \ref{eq:single}). The results are shown in Fig. \ref{fig:ratios}. Evidently, the simple analytic estimates underpredict the Monte Carlo simulation results for those isotope ratios. This is reasonable, given the discussion above, suggesting the relative abundance ratios should be in between the single event and stationary limits. At large $r,D$ (corresponding to $\tau_{\rm mix}\ll \tau_{\rm A}, \tau_{\rm B}$), the simulations also reproduce the saturation of both isotope ratios at levels comparable to those estimated from the steady state limit. The allowed values for $r, D$, thus lie in between the expectations from the two simplified limiting cases (the stationary regime and the single enrichment).

Crucially, the parameter space allowed by the two abundance ratios overlaps and is consistent with the fiducial values of $r=3\times 10^{-5}\mbox{ yr}^{-1}$ and a diffusion coefficient of $D=0.1 \mbox{ kpc}^2\mbox{ Gyr}^{-1}$.

\subsection{Number of events contributing to enrichment}
\label{sec:Numberevents}
\begin{figure}
\includegraphics[width=0.225\textwidth]{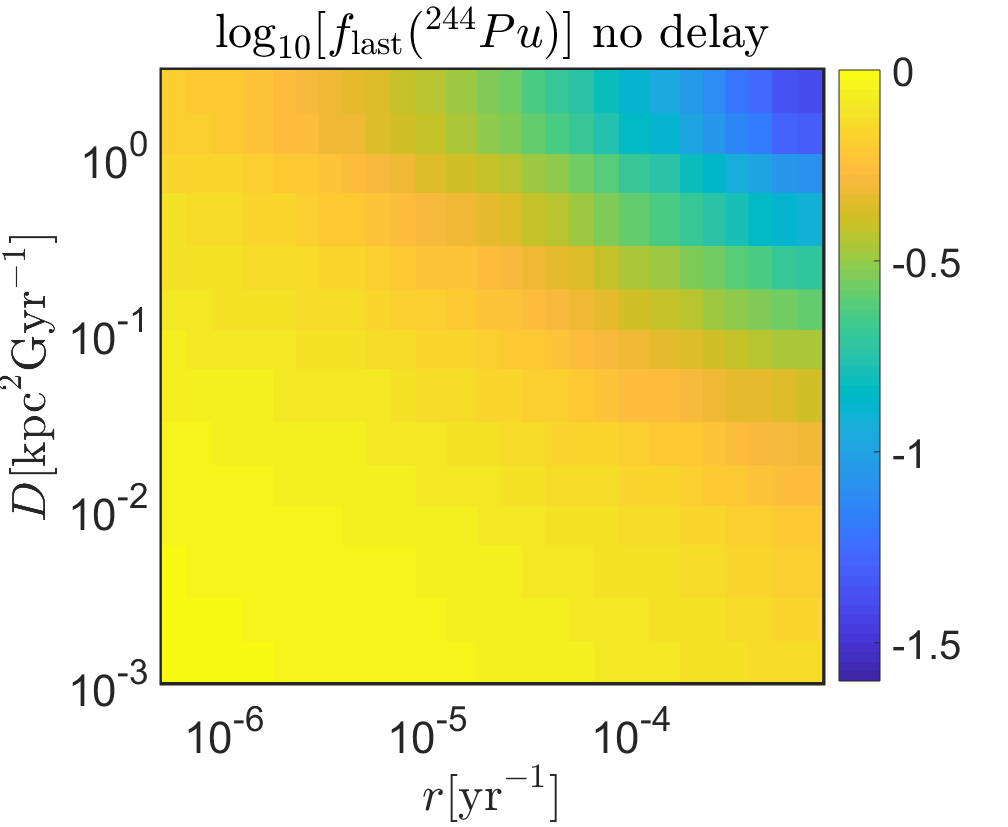}
\includegraphics[width=0.225\textwidth]{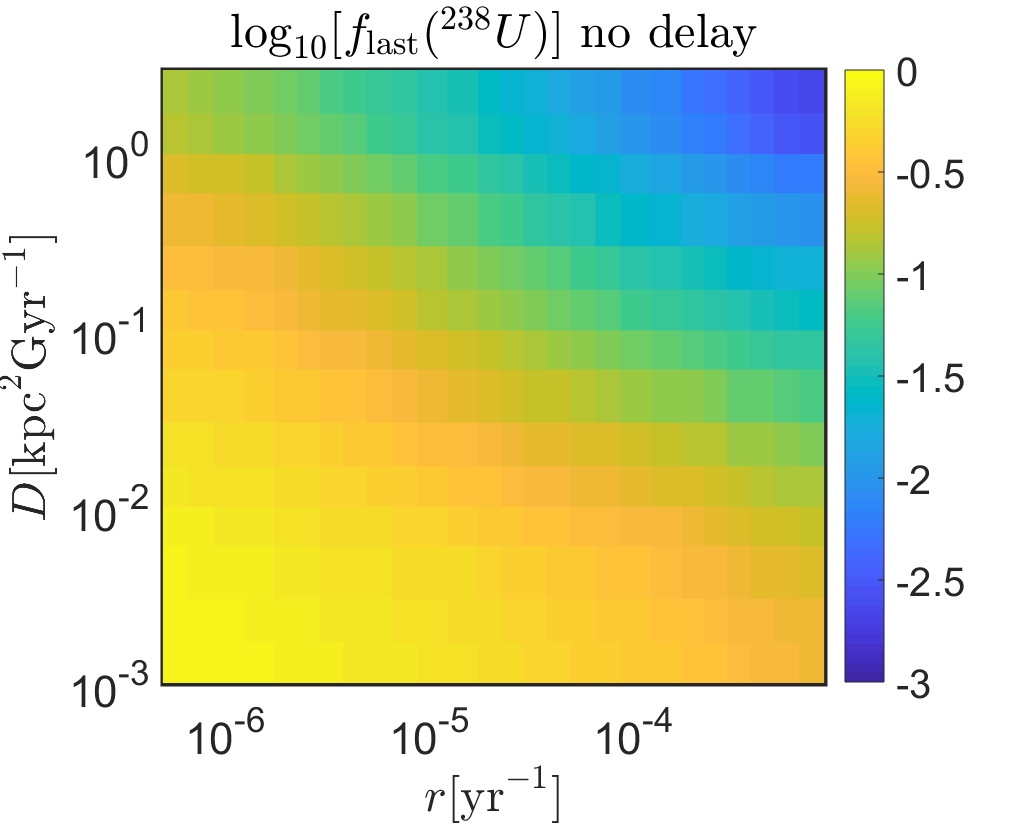}\\
\includegraphics[width=0.225\textwidth]{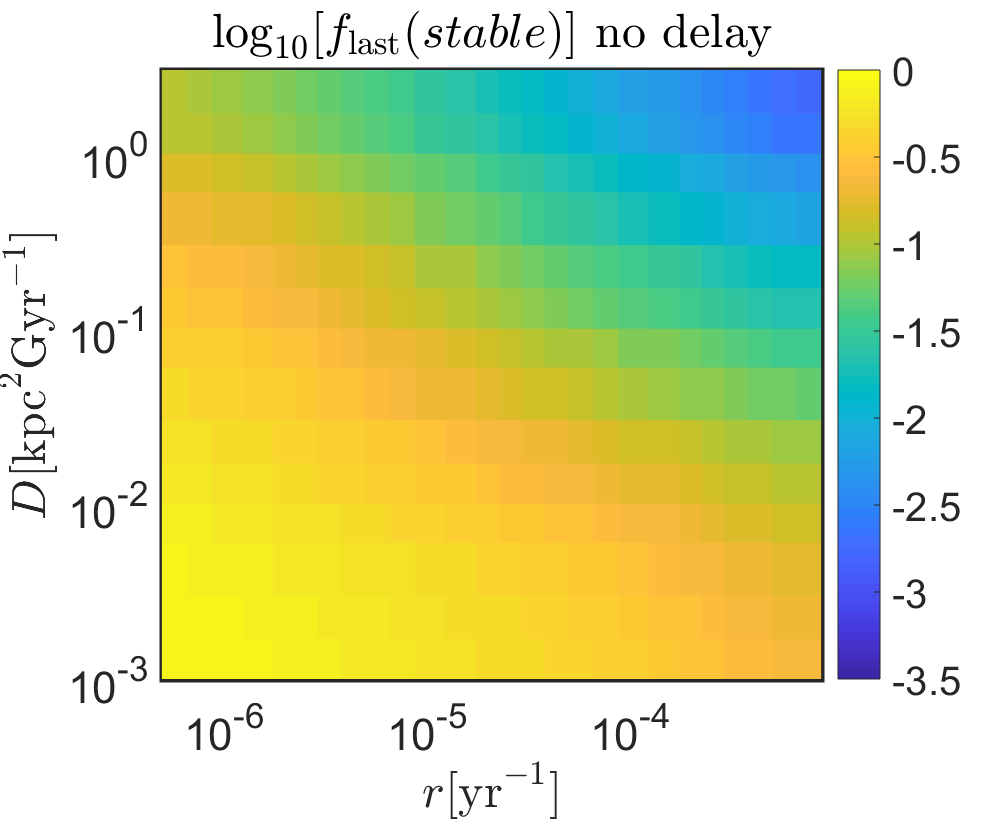}
\includegraphics[width=0.225\textwidth]{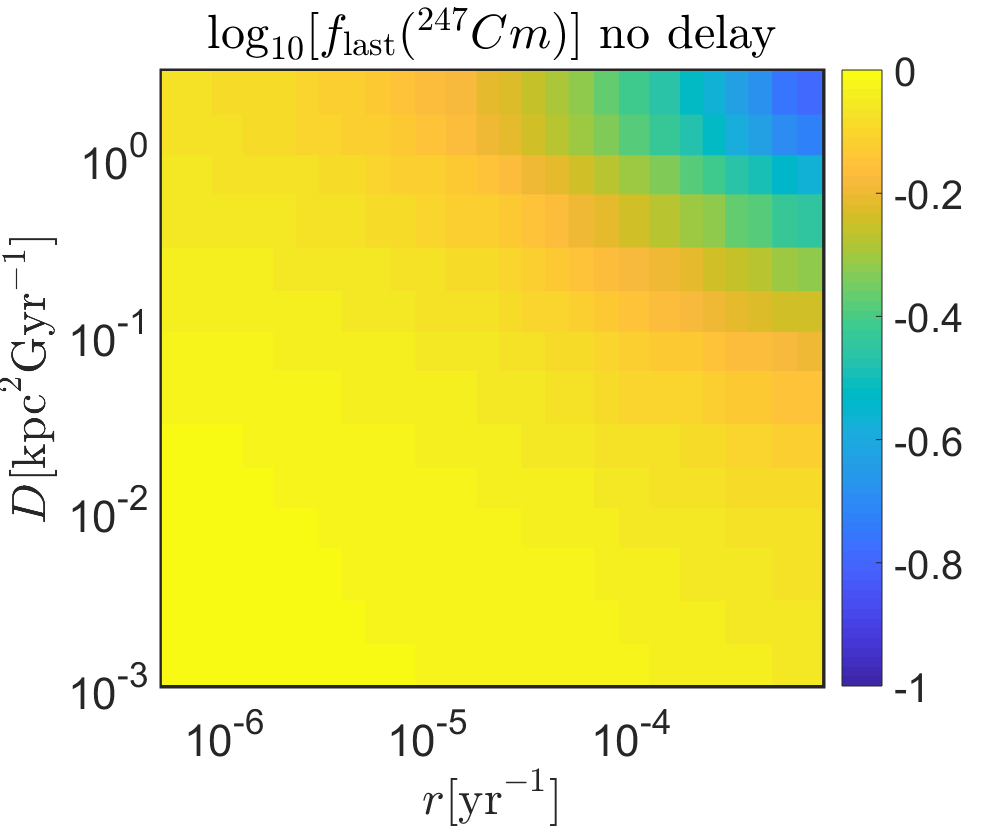}
\caption{$f_{\rm last}$ - The relative contribution to a given element's density from the most significant single event in its past as compared to its total density at the ESS. From top left in clockwise order, results are shown for $^{244}{\rm Pu}, ^{238}{\rm U}, ^{247}{\rm Cm}$ and a stable element.}\label{fig:flast}
\end{figure}

As mentioned in \S \ref{sec:ratios}, the number of past events contributing to the observed abundance of a given element is a crucial issue (see  also \citealt{cote2019ApJb}), in part, due to its usefulness in deriving simple analytic approximations. To explore this issue, we define a parameter $f_{\rm last}$, which is the relative contribution to a given element's density from the most significant single event in its past as compared to its total density at the ESS. By construction, $0\leq f_{\rm last} \leq 1$ where $f_{\rm last} \rightarrow 1$ indicates that the abundance of the element is dominated by a single event. The value of $f_{\rm last}$ for $^{244}{\rm Pu}, ^{238}{\rm U}, ^{247}{\rm Cm}$ is shown in figure \ref{fig:flast}. As expected, for elements with shorter radioactive decay times, $f_{\rm last}$ becomes larger. For our fiducial values of the model parameters,  $r=3\times 10^{-5}\mbox{ yr}^{-1}, D=0.1 \mbox{ kpc}^2\mbox{ Gyr}^{-1}$, we find that $f_{\rm last}(^{244}\mbox{Pu})\approx 0.5, f_{\rm last}(^{247}\mbox{Cm})\approx 0.8$, implying the assumption that the ESS abundance is dominated by a single event is relatively well justified for $^{247}{\rm Cm}$ and marginal for $^{244}{\rm Pu}$.
For the same model parameters, we find that $f_{\rm last}\approx 0.08$ for stable elements. Since the number of events contributing to enrichment is $\lesssim f_{\rm last}^{-1}$, this implies the abundance of a stable element at the ESS is typically dominated by more than $f_{\rm last}^{-1}\approx 10$ events. 

Applying this argument to  metal poor stars (lower values of [Fe/H]  or, equivalently, formed at earlier times than the Sun), it becomes more likely that stars have only been enriched by a single event. In particular, for a constant SFR, $\langle\mbox{[Fe/H]}\rangle=-2$ is obtained at a MW age of $\approx 5\times 10^7 \mbox{ yr}$ which for our fiducial values of $r,D$ is much shorter than $\tau_{\rm mix}$ as given by equation \ref{eq:nratio} and therefore guarantees with high probability that MW born stars with such abundances are dominated by a single event. The situation is somewhat different in case the MW SFR follows the cosmic one. In this case, the MW is at an age of $\approx 7\times 10^8\mbox{ yr}$ when it reaches $\langle\mbox{[Fe/H]}\rangle=-2$, which is comparable and even somewhat larger than $\tau_{\rm mix}$. This makes multiple contributions to a given star's enrichment more probable, even for extremely metal poor stars.

\subsection{Varying delay times, offsets and combining the different constraints}
\label{sec:varassump}

\begin{figure}
\centering
\includegraphics[width=0.4\textwidth]{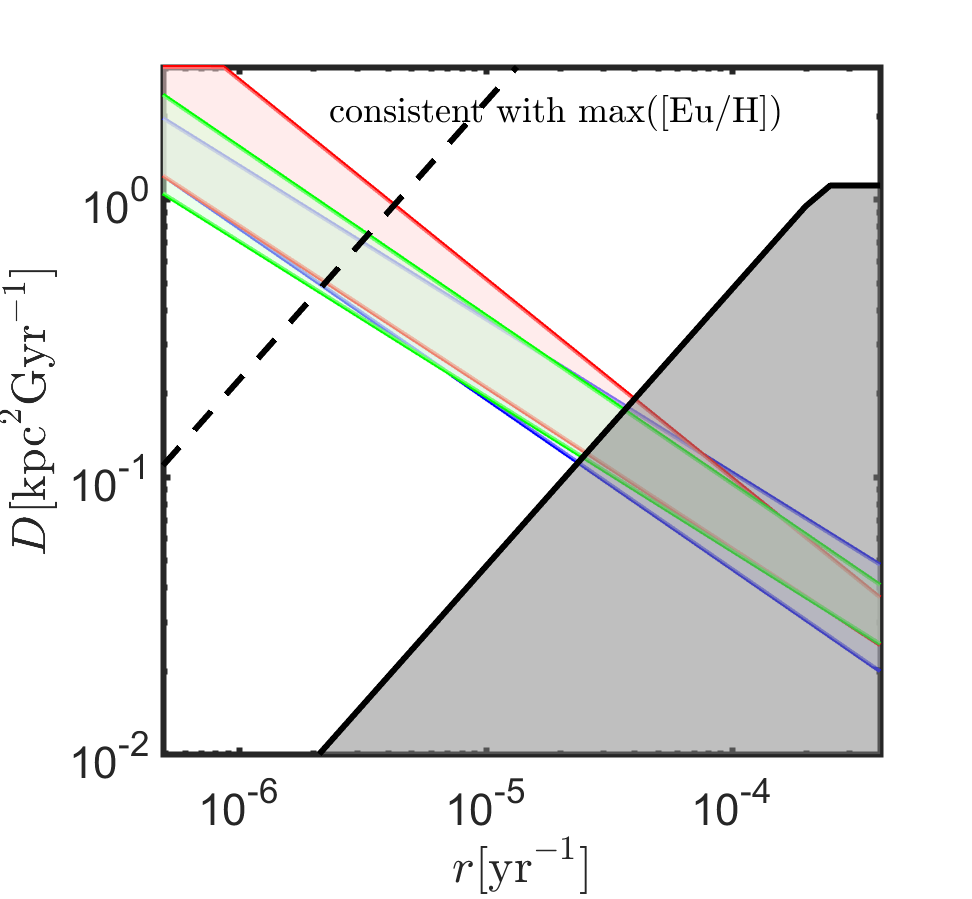}
\caption{Coloured areas not overlapping with the gray region depict the allowed range of $D,r$ given by combining the different constrained outlined in this work. A solid (dashed) black line depicts the limit from the maximum observed value of [Eu/H] assuming a Eu mass per event of $m_{\rm Eu}=10^{-4}M_{\odot}$ ($m_{\rm Eu}=10^{-3}M_{\odot}$). The blue region depicts the results for our fiducial model with short delay times and small offsets. Two additional models, with a DTD $\propto t^{-1}$ and with or without significant vertical offsets are shown in green and red regions respectively.}\label{fig:combine}
\end{figure}

The parameter space allowed by the isotope ratios of $^{244}$Pu/$^{238}$U and $^{247}$Cm/$^{238}$U as well as by the observed scatter of stable elements all overlap with each other. This is good evidence that the model presented here can self-consistently account for these independent observations and lends credence to this approach. The most constraining of these constraints is the $^{247}$Cm/$^{238}$U isotope ratio. 
Accounting in addition for the constraint from the maximum observed value of [Eu/H] (see \S \ref{sec:rare}) we can obtain further limits on the rate and the diffusion coefficient: $r\lesssim4\times 10^{-5}\mbox{ yr}^{-1}, D\gtrsim 0.1 \mbox{ kpc}^2\mbox{Gyr}^{-1}$.
The parameter space allowed by the combination of these different constraints is presented in the blue shaded region in figure \ref{fig:combine}.

The results presented thus far were calculated under the assumption of (i) short time delays between formation of the progenitors and the corresponding $r$-process events and (ii) $r$-process event locations with small offsets from the progenitor stellar population.
As mentioned in \S \ref{sec:model}, these assumptions are clearly well motivated for collapsar or other peculiar ccSNe models for $r$-process formation. However, for BNS, the situation is less clear.
The strong dependence of the delay time on the initial separation has lead many authors to consider a DTD between formation and merger of BNS that is proportional to $t^{-1}$ (e.g. \citealt{Piran1992}), while more recent investigations have found that the Galactic population of BNS requires a much steeper DTD \citep{BP2019}. Similarly, the Galactic population of BNS suggests that the majority of BNS are formed with weak kicks, corresponding to changes in the center of mass velocity of $\Delta v_{\rm cm}\lesssim 10\mbox{km s}^{-1}$ \citep{BP2016}. At such low velocities, even if the kicks are oriented completely perpendicular to the Galactic disc, they would lead to vertical oscillations with a length scale of $z_{\rm typ}\sim \Delta v_{\rm cm} P_z/2\pi\sim 0.08 \mbox{ kpc}$, where $P_z\sim 50 \mbox{ Myr}$ is the time-scale for vertical oscillations in the Galactic potential. Since $z_{\rm typ}<h_z$, this suggests that the assumption that most BNS will merge in the thin disk is motivated for such kicks. If instead the typical kicks are much larger, BNS may commonly merge at larger offsets to the Galactic plane.

To address these assumptions regarding the DTD and the offsets we repeat all the calculations presented in this work for two more models\footnote{Significant vertical offsets necessitate also long time delays, therefore we do not assume a third case with large offsets but short time delays}, (a) Same merger locations but with a DTD $\propto t^{-1}$ above $t_{\rm min}=35$Myr (the latter is taken to best fit the Galactic population, see \citealt{BP2019} for details) and (b) same DTD as in the previous case, but with merger locations corresponding to vertical harmonic oscillations with an amplitude corresponding to vertical kicks of $200\mbox{km s}^{-1}$, i.e. $z_{\rm typ}(\Delta v_{\rm cm}=200\mbox{km s}^{-1})\approx 1.6\mbox{ kpc}$.

We find that the results in the models including time delays and offsets are very close to the results presented for our fiducial model in the previous sub-sections. This is demonstrated in figure \ref{fig:combine} where we plot the allowed regions in the $r,D$ parameter space for all three models. We conclude that the results presented here are largely insensitive to these assumptions.

As mentioned in \S \ref{sec:model}, another potential source of concern is differential rotation. Since the permitted parameter space for $r,D$ found by our analysis satisfies $\tau_{\rm mix}<\tau_{\rm rot}$, the assumption that differential rotation can be neglected turns out to be well justified. In particular note the good agreement found between the allowed region of parameter space required by our analysis of the scatter in stable element abundances and that found from, e.g., the $^{247}$Cm abundance, which has a radioactive lifetime of $\tau=22.5\mbox{Myr}\ll \tau_{\rm rot}$. Its abundance could therefore not have been affected by differential rotation.

\subsection{Implications of stochastic iron production}
\label{sec:ironstochastic}
Our analysis presented thus far has explicitly taken into account the stochastic nature of $r$-process events in order to constrain both the typical levels of, and the fluctuations in, element abundances. However, we used the approximation that the iron abundance progresses smoothly over time and the stochastic nature of its production may be ignored. From a practical point of view, this approximation is necessary, due to the vast number of ccSNe in the history of the Galaxy, relative to $r$-process events ($r_{\rm cc}/r\sim 10^3$, where $r_{\rm cc}$ is the ccSNe rate). This implies an increase by a similar factor in computation time. At the same time, this large ratio between the rates is also the reason why this assumption is reasonable at late times (e.g. at the time of the solar system's formation), at which the contribution to the iron abundance from any given nearby ccSNe is much smaller than the background level of abundance contributed to by past ccSNe (whose ejecta is already well mixed). More quantitatively, this can be understood by considering equation \ref{eq:nratio} with $r_{\rm cc}\sim 0.035 \mbox{ yr}^{-1}$ \cite{Li2011}, which yields $\tau_{\rm mix,cc}\approx 10\mbox{ Myr}$. Clearly, $\tau_{\rm mix}$ is shorter than any of the radioactive lifetimes considered in this work. Furthermore, it is of the order of $t_*, T_{\rm fade}$ which are minimal time-scales for which stochasticity can be imprinted onto element abundances.
Nonetheless, at early times of the Milky Way evolution, when much fewer ccSNe have occurred in the vicinity of the solar system, the stochastic nature of both $r$-process events and ccSNe must be taken into account and can have a detectable impact on the abundances. This is the approach we adopt below.We caution however, that at early times, processes other than turbulent diffusion in the ISM may be driving the abundance fluctuations. For example, as mentioned in \S \ref{sec:scatter}, a significant fraction of metal poor stars may have originated from dwarf galaxies tidally disrupted onto the Milky Way halo at early times, weren't born in situ. Their abundance fluctuations will therefore not be dominated by turbulent diffusion. Applying the model described here to observations requires therefore a carefully selected sample of stars.

We employ the same Monte Carlo calculation outlined in \S \ref{sec:model}, but allowing for both ccSNe and $r$-process events to occur stochastically. For the former, we consider a Galactic rate of $r_{\rm cc}= 0.035 \mbox{ yr}^{-1}$. To enable computation at realistic time-scales, we consider times up to $t_{\rm max}=10^9 \mbox{ yr}$. as well as a sub-volume of the Galaxy (in which the number of events is much smaller than the Galactic total) that extends up to a distance $r_{\rm max}=1\mbox{ kpc}$ from the solar system. As long as $r_{\rm max}\gg (Dt_{\rm max})^{1/2}$ (which is the case for the values of $D$ we consider), there is not enough time for events occurring beyond this radius to contribute to local abundances. As a result, our assumption that events with a distance to the solar system further than $r_{\rm max}$ can be ignored, is well justified.

\begin{figure}
\centering
\includegraphics[width=0.3\textwidth]{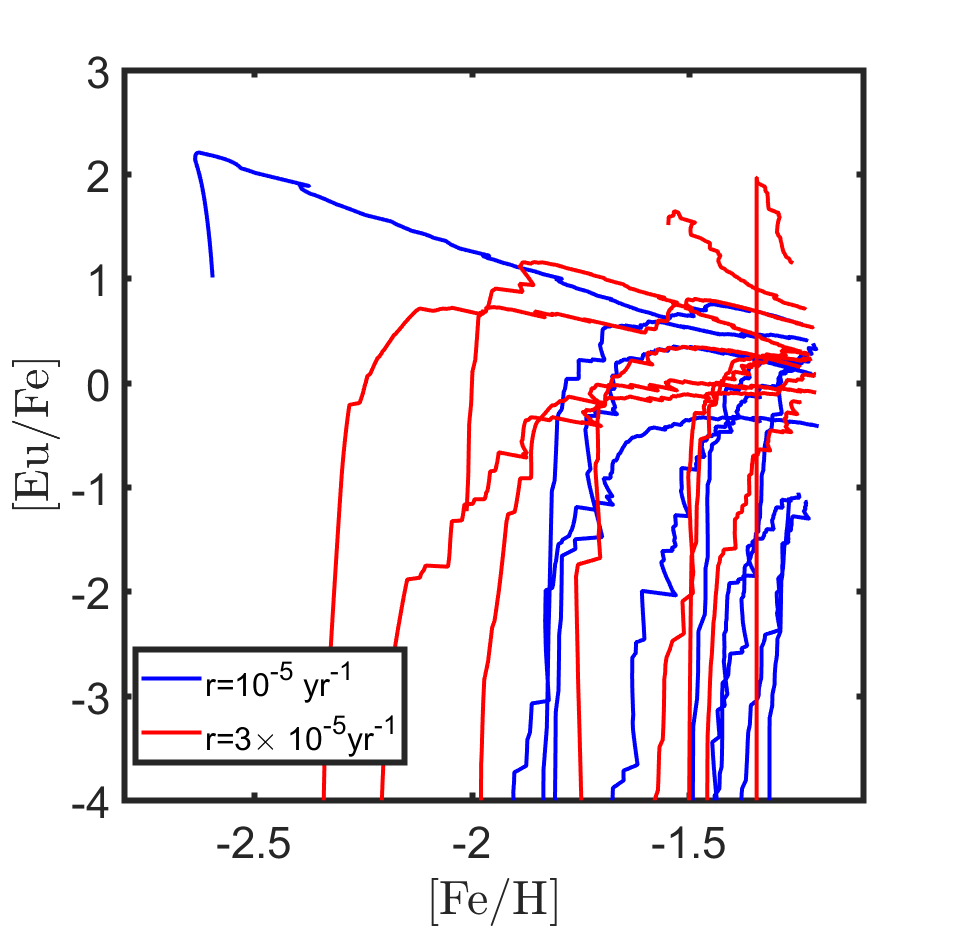}\\
\includegraphics[width=0.3\textwidth]{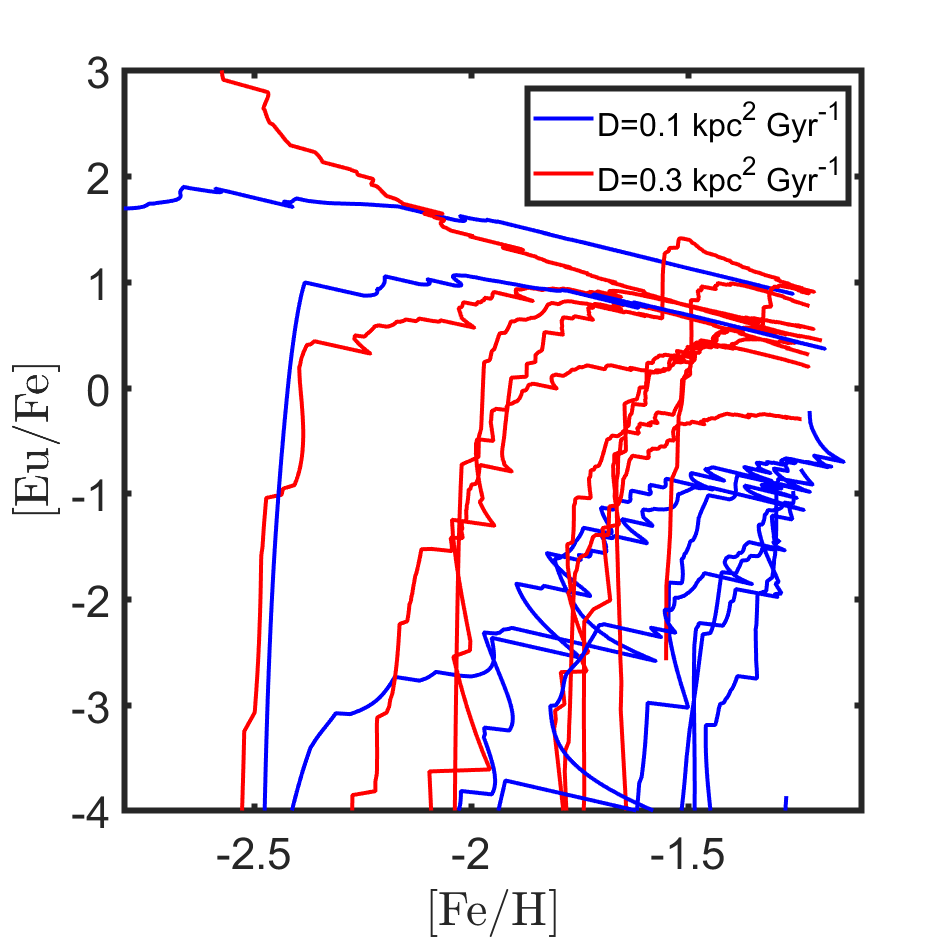}
\caption{Evolution tracks in the $\{\mbox{[Eu/Fe], [Fe/H]}\}$ plane for different realizations of our Monte Carlo simulation accounting for the stochasticity of both $r$-process events and ccSNe. Top: Same diffusion coefficient ($D=0.3 \mbox{ kpc}^{2}\mbox{ Gyr}^{-1}$) and different rates. Bottom: Same $r$-process rate ($r=3\times 10^{-5} \mbox{ yr}^{-1}$) and different diffusion coefficient. }\label{fig:tracks}
\end{figure}

In figure \ref{fig:tracks} we present different realizations of our Monte Carlo calculation in the [Eu/Fe], [Fe/H] plane for different diffusion coefficients and different $r$-process rates.
Similar to our results with stochastic $r$-process events only (see \S \ref{sec:scatter}) we find that lower values of $D, r$, cause a broadening of the predicted abundance patterns, as well as an earlier rise of [Eu/Fe] as a function of [Fe/H]. This can be estimated by considering equation \ref{eq:nratio}. To a first approximation, the lowest [Fe/H] for which significant $r$-process rich stars should be observed is such that $\langle t(\mbox{[Fe/H]})\rangle\approx \tau_{\rm mix}$ where $\langle t(\mbox{[Fe/H]})\rangle$ is the typical time it takes to reach an iron abundance of [Fe/H] for a given star formation and ccSNe rate. Since there are multiple $r$-process stars with strong $r$-process enrichment and $\mbox{[Fe/H]}<-2$, we can take $\langle t(\mbox{[Fe/H]=-2})\rangle$ as a characteristic value. The later strongly depends on the assumed star formation at the early stages of the MW evolution (see discussion in \S \ref{sec:Numberevents}). Generally, an SFR following the cosmic rate, leads to larger values of $\langle t(\mbox{[Fe/H]})\rangle$ and is therefore more easy to reconcile with the implied rate and diffusion coefficient found in this work. We consider this star formation rate throughout this section.

\begin{figure*}
\centering
\includegraphics[width=0.3\textwidth]{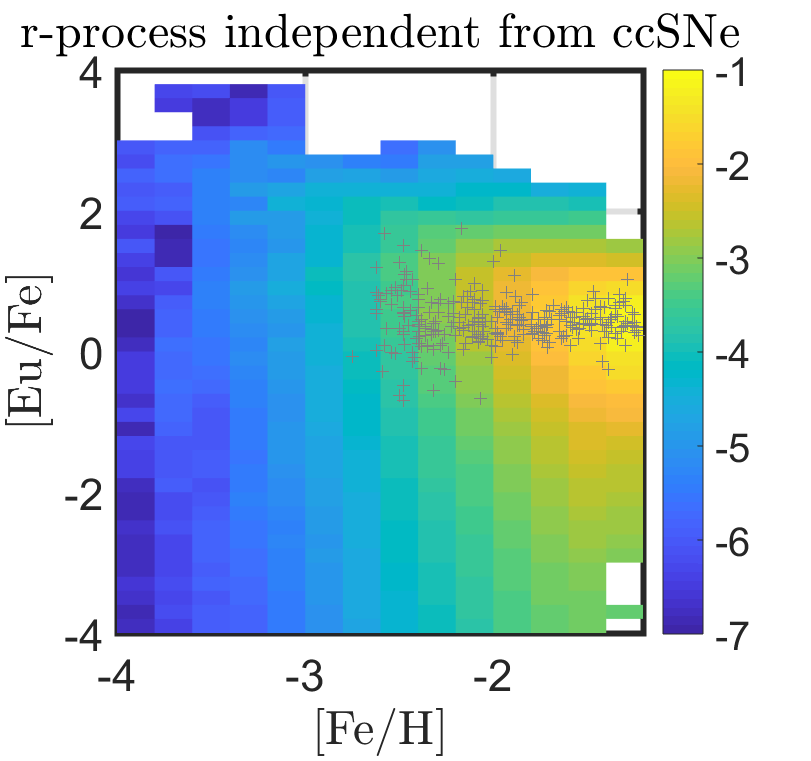}
\includegraphics[width=0.3\textwidth]{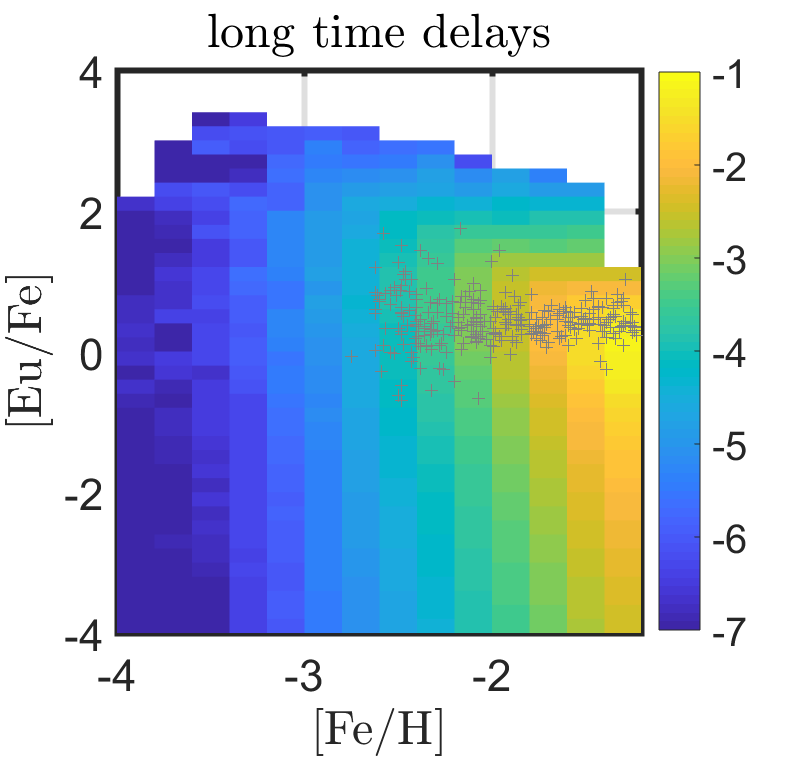}
\includegraphics[width=0.3\textwidth]{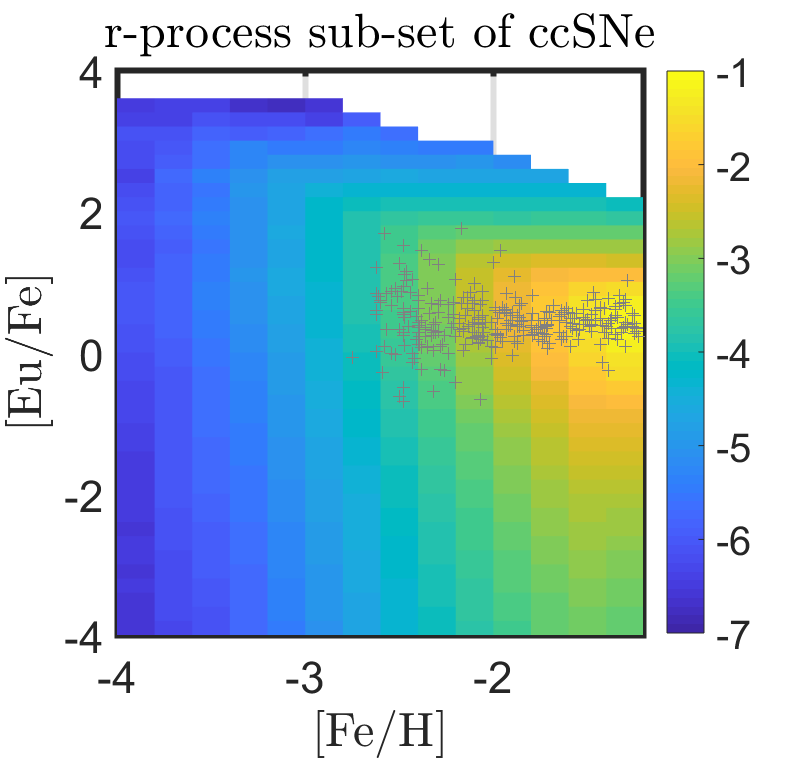}\\
\includegraphics[width=0.3\textwidth]{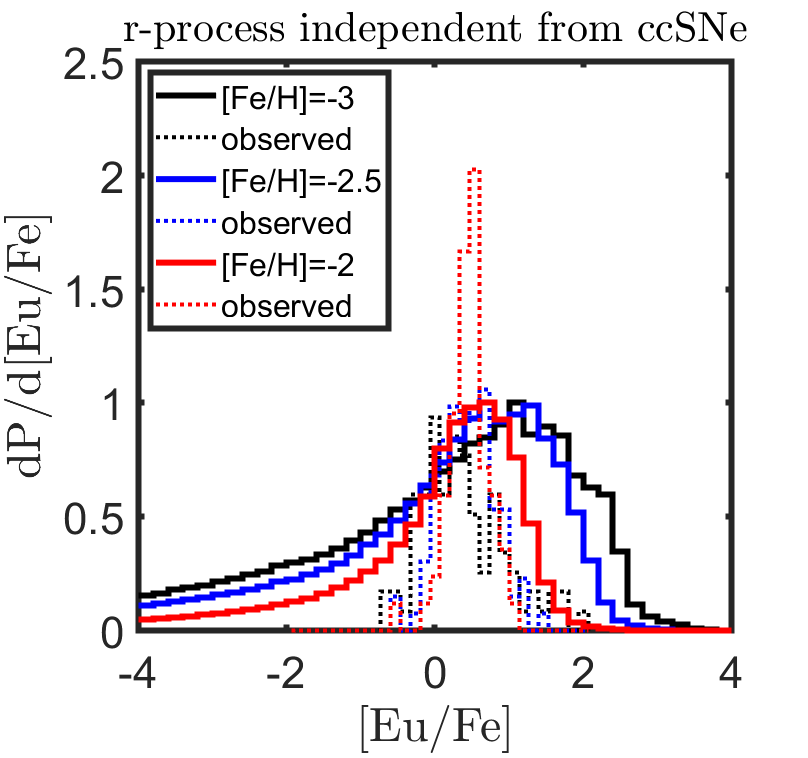}
\includegraphics[width=0.3\textwidth]{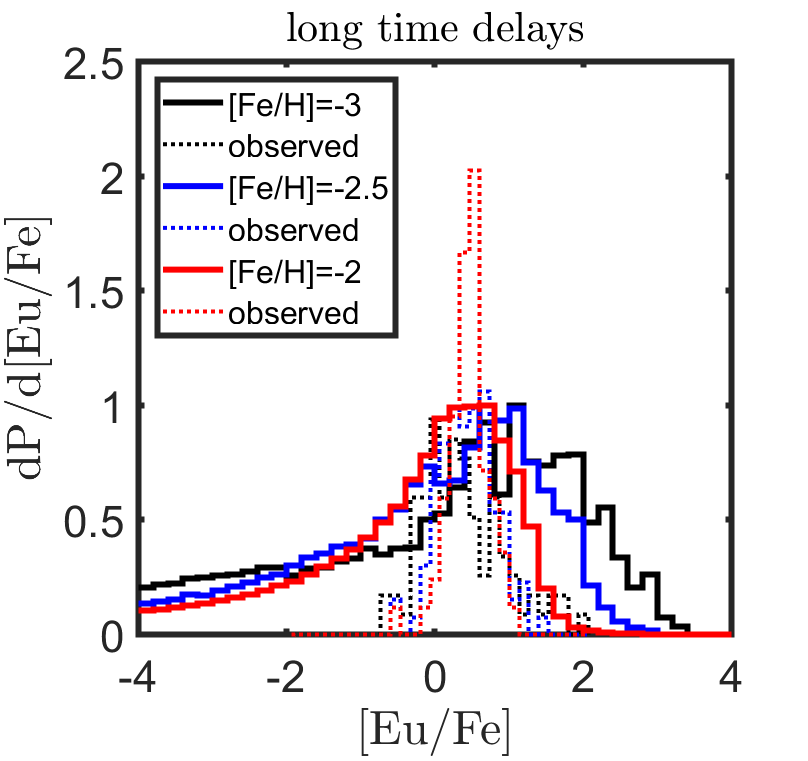}
\includegraphics[width=0.3\textwidth]{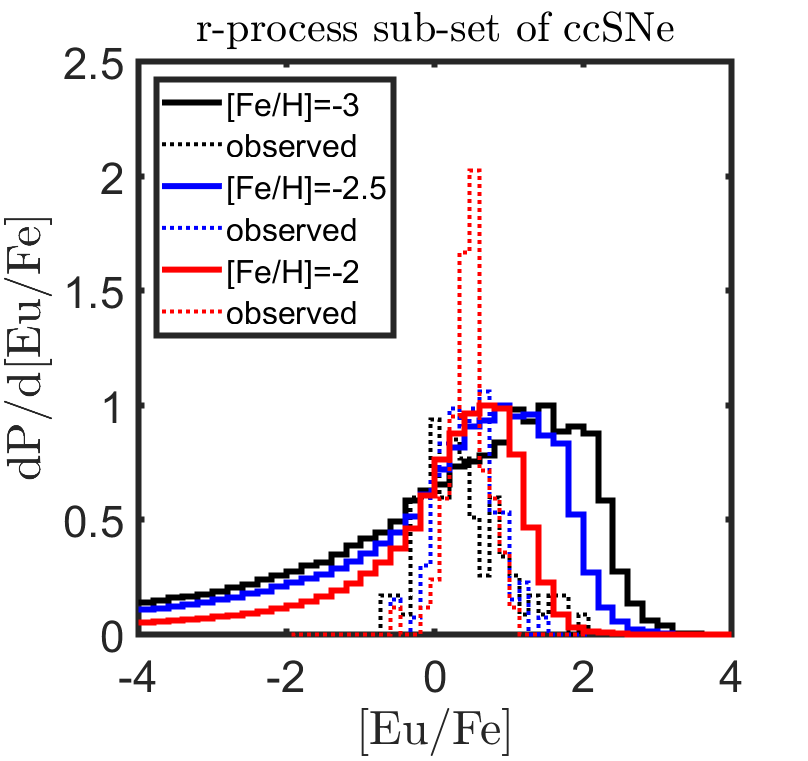}
\caption{Top: 2D probability maps of $\{\mbox{[Eu/Fe], [Fe/H]}\}$ in metal poor stars arising from three different models (see \S \ref{sec:ironstochastic} for a full description) taking into account stochasticity of both $r$-process events and ccSNe and assuming $r=3\times 10^{-5}\mbox{ yr}^{-1}, D=0.3 \mbox{ kpc}^2\mbox{ Gyr}^{-1}$. Colours denote the base ten logarithm of the normalized 2D probability density function. Over-plotted are stellar abundances from the SAGA database. Bottom: 1D projections of the probability distributions for fixed values of [Fe/H] and varying [Eu/Fe].}\label{fig:2D}
\end{figure*}

In figure \ref{fig:2D} we present also the 2D distributions of metal poor stars (for which the stochastic nature of iron production can have an observable impact) in the $\{\mbox{[Eu/Fe], [Fe/H]}\}$ plane resulting from $2\times 10^4$ realizations of our Monte Carlo simulations in which both ccSNe and $r$-process events are stochastic. We have considered three limiting options to take into account different possibilities regarding the nature of the $r$-process events. These options are deliberately chosen to probe the edges of the expected parameter space, in order to better demonstrate the effect of the underlying model parameters on the result.
\begin{enumerate}
    \item No time delay between star formation and $r$-process events, but with $r$-process events occurring at different locations and times than ccSNe. This mimics the case in which BNS mergers dominate the $r$-process, but they are dominated by systems with short delays between BNS formation and merger.
    \item A time delay between star formation and $r$-process events following a DTD$\propto t^{-1}$ above $t_{\rm min}=35\mbox{ Myr}$, and $r$-process events occurring at different locations and times than ccSNe. This mimics the case of BNS mergers with longer time delays.
    \item No time delay between star formation and $r$-process events, and with $r$-process events being a sub-set of ccSNe, i.e. sharing the same locations and times as a fraction (determined by the rate) of ccSNe. This mimics the case in which $r$-process production is dominated by same rare type of ccSNe, e.g. collapsars.
\end{enumerate}
We find that for a fixed $r,D$ and star formation rate, cases (1), (3) result in very similar distributions to each other. These distributions generally result in a larger scatter than seen in observed metal poor stars. One possibility is that this difference is due to the observational bias against measuring lower $r$-process abundances in metal poor stars. Case (2) leads to a longer delay between star formation and the first $r$-process events, and therefore results in a distribution that is skewed towards slightly larger values of [Fe/H]. Nonetheless, even case (2) that has quite long time delays, cannot be ruled out by current observations.

\section{Conclusions}
We have studied the effects of turbulent gas diffusion on the observed $r$-process abundances in Milky Way stars, by a combination of an analytical approach and a Monte Carlo simulation. This modelling enables us to take advantage of not just the mean $r$-process abundance evolution observed across Galactic stars of different metallicities, but also to utilize higher modes of the observed distribution in order to infer the Galactic rate of $r$-process events and the coefficient for turbulent gas diffusion in the Galaxy. Higher $r$-process event rates and faster diffusion times, lead to more efficient mixing of the $r$-process material which in turn leads to a reduced scatter of $r$-process abundances and causes $r$-process enriched stars to start appearing  at lower metallicity values.
We use three independent observations to constrain the model parameters: (i) the scatter of radioactively stable $r$-process elements at a fixed value of [Fe/H], (ii) The largest $r$-process enrichment values observed in any of the observed stars and (iii) the isotope abundance ratios of different radioactive $r$-process elements ($^{244}$Pu/$^{238}$U and $^{247}$Cm/$^{238}$U) at the ESS as compared to their formation ratios.

Our results indicate that the Galactic $r$-process rate and the diffusion coefficient are respectively $r\lesssim 4\times 10^{-5}\mbox{ yr}^{-1}, D\gtrsim 0.1 \mbox{ kpc}^2\mbox{Gyr}^{-1}$ with the additional requirement that below those limiting values, a rough correlation holds between the required values of $r,D$ such that $D\approx 0.3 (r/10^{-5}\mbox{ yr}^{-1})^{-2/3}\mbox{ kpc}^2\mbox{ Gyr}^{-1}$. This implies that the time between two $r$-process events that significantly enrich the same location in the MW, is of the order of $\tau_{\rm mix}\approx 100-200\mbox{ Myr}$. This in turn means that a fraction of $\sim 0.8$ ($\sim 0.5$) of the ESS $^{247}$Cm ($^{244}$Pu) abundance is dominated by one $r$-process event in its past. At the same time, for radioactively stable elements, their ESS abundance is dominated by contributions from $\sim 10$ different events in their past. 
For metal poor stars (with [Fe/H]$\lesssim -2$), their $r$-process abundances are expected to be dominated by a single event, if the MW SFR is assumed to be constant in time. If instead, the MW SFR follows the Cosmic one, this however is no longer the case. The spatial abundance variations due to turbulent mixing are shown to increase linearly with distance for $r\lesssim 1$kpc with a normalization that decreases with larger values of $r$ and / or $D$ in an independent way than probed by the other considerations mentioned above. Measuring such spatial variations could therefore provide an additional observational test for the diffusion coefficient and the rate.

The upper limit on the rate of $r$-process events found in this work, is consistent with independent estimates based on $r$-process abundances of ultra-faint and classical dwarfs, which at a 2$\sigma$ level lead to $6\times 10^{-6}\mbox{ yr}^{-1}-4\times 10^{-5}\mbox{ yr}^{-1}$ \citep{Beniamini2016}. In particular, the lower bound in this range, combined with the correlation between the required values of $r$ and $D$ found in the present work, can be translated to an upper limit on the diffusion coefficient, $D\lesssim 1\mbox{ kpc}^2\mbox{ Gyr}^{-1}$.

Our limits on the $r$-process rate and the diffusion coefficient become more constraining for larger $r$-process masses produced per event. For example, increasing the Eu mass produced per event from $10^{-4}M_{\odot}$ to $10^{-3}M_{\odot}$ (representative of the expected difference between the mass created per event in BNS mergers and collapsars, see \citealt{Siegel2019}) leads to an upper limit on the rate that is decreased by a factor of ten $r\lesssim 4\times 10^{-6}\mbox{ yr}^{-1}$ and to a lower limit on the diffusion coefficient that is increased by a factor of five $D\gtrsim 0.5\mbox{ kpc}^2\mbox{Gyr}^{-1}$ (see figure \ref{fig:combine}). This is driven by the constrain from the maximally enriched stars observed in the solar neighborhood.
As apparent from the discussion above, this leaves a rather narrow allowed parameter space for collapsar models in the $r,D$ parameter space.

The results regarding the permitted range of $r,D$ found in this work are rather insensitive to the assumptions on the delay times between star formation and $r$-process enrichment and on the assumed locations of the $r$-process events within the Galaxy. In addition, we have verified that the stochastic nature of the iron production can be reasonably ignored on the timescales relevant to the mixing of $r$-process elements. That being said, if future observations continue to reveal highly $r$-process enriched stars at very low metallicities ([Fe/H]$\lesssim -2.5$), this could potentially limit $r$-process formation channels involving long time delays between star formation and $r$-process formation. 
However, such an observation could also be ascribed to lowered star formation efficiency at the early stages of the Milky Way (e.g. \citealt{ishimaru2015ApJ,Ojima2018}) or to a different channel of $r$-process formation becoming dominant at those times, such as e.g. black hole - neutron star mergers \citep{Korobkin2012}.
A carefully selected unbiased sample of metal poor stars, whose place of birth can be confidently placed within the Galaxy would be paramount to using the detailed comparisons with observations to infer the finer details of $r$-process formation rates, locations and their turbulent mixing. 

Throughout this paper, we have assumed that metals are strongly coupled to the gas, such that turbulent mixing becomes important relatively early on, after $T_{\rm fade}\sim 5$Myr (which marks the end of the blast wave phase). This is expected to be the case for reasons that we expand on below. First, consider neutron star mergers. In these environments, the expansion velocity is so fast that the ejecta density at the condensation temperature is too low to form dust grains \citep{Takami2014}, and the blast wave dynamics are then fully appropriate for studying the early dynamics of those elements, and the time before they mix with ISM gas. 
Alternatively, for core collapse supernovae, it is well known that dust grains are formed. A major uncertainty here is whether or not dust grains survive and if they do, whether they are coupled with the gas motion.
The destruction of dust can occur through the passage of the reverse shock of the supernova remnant. Several authors have found that this process could result in the destruction of the majority of dust grains created in SNe \citep{2016A&A...589A.132B,2016A&A...590A..65M}. Furthermore, recently, \cite{Fry2018} studied the coupling of dust grains and gas under the existence of magnetic field. These authors have found that the dust grains are trapped in the shocked ISM. In this case, our assumption regarding the coupling to the gas is indeed valid. However, to test this further, we suggest that more efforts on this topic are necessary, e.g., varying the magnetic field strength and structure as well as different assumptions on the properties of dust grains formed in supernovae.

Our method is applicable to studying the abundance of various radioactive elements that existed in the ESS (e.g., \citealt{Wasserburg2006NuPhA,Fry2016ApJ,Lugaro2018}). In a future work we will also examine possible applications to star-to-star elemental variations of dwarf galaxies and clusters (see e.g. \citealt{Kirby2020} for a recent study of the star-to-star scatter in the globular cluster M15). The former will enable us to reveal the production sites of different isotopes \citep{cote2019ApJb}. The latter will provide better understanding of the formation and evolution of galaxies and clusters under different environments. 
Finally, we note that the map of the Galactic diffuse $\gamma$-ray line emission of $^{26}$Al may also be useful to reveal the diffusion coefficient of the ISM (e.g.\,\citealt{Bouchet2015ApJ,Fujimoto2018MNRAS,Wang2020ApJ}).
We will extend this work to these topics in future.

\section*{Acknowledgments}
We thank the anonymous referee for their constructive report. PB thanks Sterl Phinney, Enrico Ramirez-Ruiz, Evan Kirby, Tony Piro and Wenbin Lu for helpful discussions.
The research of PB was funded by the Gordon and Betty Moore Foundation through Grant GBMF5076.


\label{lastpage}
\end{document}